\documentclass[12pt]{article}
\usepackage{graphicx}
\usepackage{amsbsy, amstext, amssymb, amsmath,bm,bbm,wasysym}
\usepackage{algorithmic,graphicx,booktabs,hyperref,rotating,enumitem,diagbox}
\usepackage{array,stmaryrd}
\usepackage{tabularx}
\usepackage{pdflscape}
\usepackage{adjustbox}
\usepackage{subfig}
\usepackage{longtable}
\usepackage{listings}
\usepackage{multirow}
\usepackage{xr}
\usepackage{blkarray}
\usepackage{setspace}
\usepackage{mathtools}

\usepackage[natbibapa]{apacite}
\usepackage{etoolbox}
\patchcmd{\APACjournalVolNumPages}{\unskip({#3})}{}{}{} 
\patchcmd{\APACjournalVolNumPages}{\Bem{#2}}{#2}{}{}

\AtBeginEnvironment{APACrefURL}{\renewcommand{\url}[1]{}}
\AtBeginDocument{}
\AtBeginEnvironment{longtable}{\singlespacing}

\usepackage{listings}
\renewcommand{\baselinestretch}{1.2}
\usepackage{tikz}
\usetikzlibrary{shapes.geometric, arrows}
\usetikzlibrary{arrows.meta}
\usetikzlibrary{shapes,backgrounds,calc,positioning}

\newcommand{\E}{\mathop{\mathbb{E}}} 

\newcommand{\Var}{\mathrm{Var}}

\newcommand{\Vbf}{{\bm V}}

\newcommand{\Xbf}{{\bm X}}

\newcommand{\Zbf}{{\bm Z}}

\newcommand{\dbf}{{\bm d}}

\newcommand{\betabf}{\boldsymbol{\beta}}

\newcommand{\etabf}{\boldsymbol{\eta}}

\newcommand{\gammabf}{\boldsymbol{\gamma}}

\newcommand{\blind}{0}

\begin{document}

\def\spacingset#1{\renewcommand{\baselinestretch}%
{#1}\small\normalsize} 

\if0\blind
{
  \title{\bf FLAME: A Model for Duration-Dependent Risk Accumulation in Episodic Temporal Exposures}
  \author{Xinkai Zhou$^{1, *}$,
Lee Goeddel$^{2}$, Nauder Faraday$^{2}$, Ciprian M. Crainiceanu$^{3}$\\
\small{$^{1}$Department of Statistics and Data Science, Beijing Normal - Hong Kong Baptist University} \\
\small{$^{2}$Department of Anesthesiology and Critical Care Medicine, Johns Hopkins University}\\
\small{$^{3}$Department of Biostatistics, Johns Hopkins University}
}
    \date{}
  \maketitle
} \fi

\if1\blind
{
  \bigskip
  \bigskip
  \bigskip
  \begin{center}
    {\LARGE\bf FLAME: A Model for Duration-Dependent Risk Accumulation in Episodic Temporal Exposures}
\end{center}
  \medskip
} \fi

\bigskip
\spacingset{1.5}
\begin{abstract}
Emerging technologies enable continuous monitoring of temporal exposures to disease risk factors, leading to complex exposure processes characterized by subject-specific numbers and durations of exposure episodes. A key scientific question is how the number and duration of such episodes influence disease risk. Existing methods typically rely on scalar summaries or time-indexed representations and are not naturally suited to model duration-dependent risk accumulation at the episode level. We introduce the FLexible Accumulation ModEl (FLAME), a semiparametric model for risk accumulation at the level of individual exposure episodes, with duration as the primary driver of risk.  FLAME is motivated by and applied to quantifying the association between the duration of intraoperative hypotension and acute kidney injury (AKI) following cardiac surgery. The estimated risk accumulation function reveals that, although 60 one-minute hypotensive episodes are associated with an AKI probability of 0.24, a single sustained 60-minute episode increases that probability to 0.33, representing a 38\% increase despite identical total duration. These findings provide actionable insights for intraoperative hemodynamic management and demonstrate the importance of accounting for episodic exposure patterns. While motivated by cardiac surgery, FLAME is broadly applicable to other settings involving high-resolution temporal exposures. An \texttt{R} package, \texttt{flameRisk}, is provided to facilitate application of the method in practice.
\end{abstract}

%
\noindent%
\vfill

\newpage
\spacingset{2} 

\section{Introduction}\label{sec:intro}
Chronic and acute health conditions often arise from temporal exposures, including physiological, behavioral, and environmental risk factors. Understanding how risk accumulates over such exposures is critical for targeted interventions and personalized treatment. In cardiac surgery, for example, high-resolution hemodynamic measurements such as mean arterial pressure (MAP) and central venous pressure (CVP) are continuously recorded and have been shown to be associated with adverse postoperative outcomes, including acute kidney injury (AKI) \citep{goeddel2024mapcvp}. However, these continuous measurements often give rise to exposure patterns characterized by repeated transitions into and out of clinically relevant states (e.g., hypotension), resulting in irregular, discrete episodes that vary in duration. Such heterogeneity in episodic exposure patterns is not easily captured by simple summary measures of exposure.

\begin{figure}[!tbh]
\centering
\includegraphics[width=\textwidth]{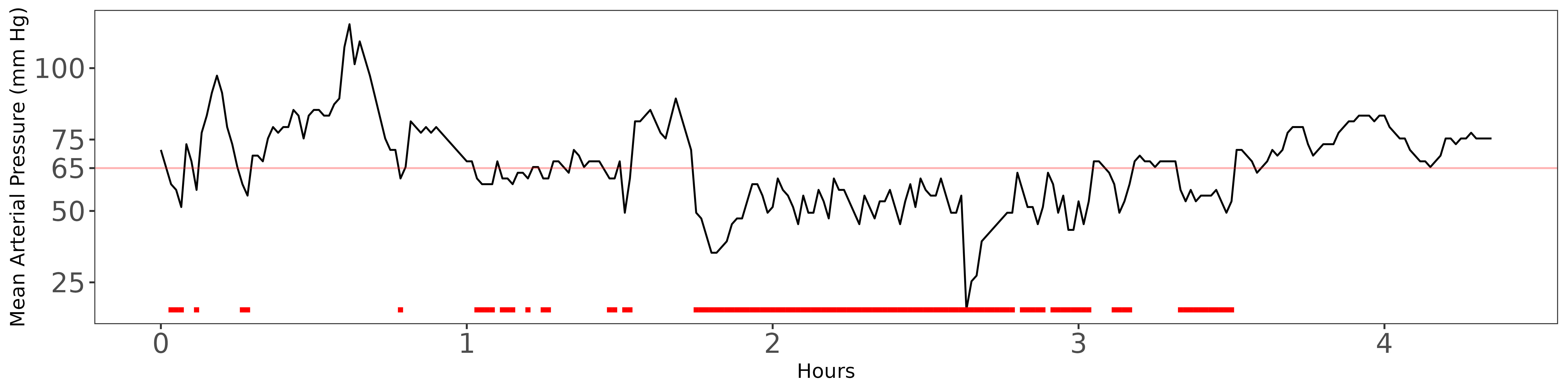}
  \caption{Mean arterial pressure (MAP) for a patient during cardiac surgery. The red horizontal line denotes the clinical threshold for hypotension (65 mm Hg). Red bars at the bottom indicate hypotensive episodes, defined as periods with MAP $\leq 65$ mm Hg.}
  \label{fig:figure1}
\end{figure}

Figure~\ref{fig:figure1} illustrates this phenomenon using MAP measurements from a patient undergoing coronary artery bypass grafting (CABG). The x-axis represents time from the start of surgery, and the red horizontal line at 65 mm Hg denotes the clinical threshold for hypotension; values below this threshold correspond to hypotensive states. The red bars at the bottom of the figure indicate the occurrence and duration of hypotension episodes. Two key features emerge. First, the patient repeatedly transitions into and out of the hypotensive state, resulting in multiple distinct episodes. Second, these episodes vary substantially in duration, ranging from minutes to hours.

This motivates the following scientific question: \textit{how do the number and duration of hypotensive episodes influence the risk of adverse outcomes (e.g., AKI after surgery)?} A natural starting point is to summarize exposure using total hypotension duration. However, such a summary cannot distinguish between patients who accumulate the same total hypotension duration through different exposure patterns. For example, two patients may each experience 60 minutes of hypotension, yet one accumulates this exposure through $60$ isolated one-minute episodes, while the other experiences a single uninterrupted 60-minute episode. Although these scenarios have identical total duration, they carry different clinical implications. In particular, risk of organ injury may increase nonlinearly with the duration of continuous exposure, and clinical interventions to manage hypotension could be more precisely administered over time based on hypotension duration.

Despite its importance, this question is not directly addressed by existing analytic approaches. Existing methods typically summarize exposure using scalar metrics, model time-indexed trajectories, or focus on lagged effects, and are therefore not well suited to capturing duration-dependent risk accumulation arising from irregular exposure episodes.

To address this gap, we propose the FLexible Accumulation ModEl (FLAME), which explicitly models risk accumulation at the level of individual exposure episodes, with duration as the primary driver of risk. Our key contribution is a statistical model that directly characterizes how risk depends on the number and duration of episodes, providing an interpretable and principled approach to analyzing episodic temporal exposures. While motivated by perioperative hemodynamic data, the model is broadly applicable to other settings involving high-resolution temporal exposures.

The remainder of the paper is organized as follows. Section~\ref{sec:literature} reviews existing methods for analyzing temporal exposures. Section~\ref{sec:FLAME} introduces the FLAME model and its implementation. Section~\ref{sec:simulation} presents simulation studies evaluating performance under diverse scenarios. Section~\ref{sec:real-data} applies FLAME to intraoperative hemodynamic data from Johns Hopkins Hospital, yielding new insights into AKI risk accumulation. Section~\ref{sec:discussion} concludes with a discussion of implications and future research directions.

\section{Literature Review}\label{sec:literature}
Temporal exposures are the de facto reality in many research domains, including pharmacoepidemiology \citep{rothman2008modernEpi}, environmental health \citep{Nieuwenhuijsen2003Exposure}, and cardiac surgery \citep{goeddel2024mapcvp}. Historically, limitations in measurements and a need for interpretable models gave rise to a rich literature using summaries of temporal exposures. For example, the weighted cumulative exposure (WCE) approach \citep{breslow1983multiplicative, thomas1988models} aggregates past exposures using a weighted sum, where the weights are either predetermined \citep{vacek1997assessing, berry1979asbestosis} or estimated \citep{hauptmann2000analysis}. In cohort studies, WCE can be calculated at each follow-up time and then used in a Cox model as a time-dependent covariate \citep{sylvestre2009flexible, xiao2014flexible}. Generalized additive models (GAMs) allow for flexible modeling of scalar exposures \citep{hastie1986gam, Crainiceanu2005WinBUGS,ruppert2003semiparametric, wood2017gam, umlauf2018bamlss} and recent advances allow many scalar exposures to be flexibly modeled together \citep{Hu2023Exposome,lin2006component, Panagiotelis2008FunctionSelection, ravikumar2009SAM, marra2011practical, Scheipl2012STAR, Bai2022spikeGAM}.

However, advances in data collection have changed both the structure of exposure data and the scientific questions of interest. Modern monitoring technologies generate data at much higher temporal resolutions, raising the question of whether scalar summaries are sufficient to capture complex exposure patterns. This has motivated developments in functional data analysis \citep{ramsaysilv2005, crainiceanu2023book}, which model the entire exposure trajectory. In particular, scalar-on-function regression (SOFR) \citep{tikhonov1963regularization, wahba1990spline, reiss2017sofr} and functional additive models \citep{muller2008fam, mclean2014functional} have been developed for exposures defined on a common domain and the same interpretation of data at the same time point. Despite their flexibility, these approaches do not directly target risk accumulation over irregular, discrete exposure episodes that vary in number and duration across subjects. Perhaps the most closely related existing framework is that of distributed lag models (DLMs) \citep{almon1965distributed, schwartz2000distributed, zanobetti2000generalized}, which we review in detail and contrast with our method in Section~\ref{sec:relation}.

\section{Methods}\label{sec:methods}
The data structure in this paper is the one suggested by the cardiac surgery example introduced in Section~\ref{sec:intro}. Let $Y_i$ for $i=1,\ldots,I$ be a scalar outcome of interest for subject $i$ (e.g., AKI status) with an exponential family distribution and $\Xbf_i = (X_{i1}, X_{i2}, \ldots, X_{ip})^t$ be a $p$-dimensional vector of baseline covariates (e.g., age and sex).  Denote by $J_i$ the number of exposure episodes and by  $\boldsymbol{Z}_i=(Z_{i1}, \ldots,Z_{iJ_i})^t$ the vector of durations of exposure episodes. For example, in Figure~\ref{fig:figure1}, there were $J_i=15$ episodes; the first episode lasted for $Z_{i1}=3$ minutes and the longest episode lasted for $Z_{i11}=63$ minutes. The observed data are of the form $(Y_i, \Xbf_{i}, \boldsymbol{Z}_i)$. 

\subsection{Flexible Accumulation Model}\label{sec:FLAME}
Our key modeling insight is that each episode contributes risk as an unknown function of its duration. Let $f(\cdot)$ be this function, which will be referred to as the \emph{risk accumulation function} (RAF). We propose the following FLexible Accumulation ModEl (FLAME):
\begin{equation}
    g\{\E[Y_i]\} = \Xbf_i\betabf + \sum_{j=1}^{J_i} f(Z_{ij}),
    \label{eqn:FLAME}
\end{equation}
where $g(\cdot)$ is a canonical link function, $\Xbf_i\betabf$ accounts for the risk associated with the person-specific covariates, and $\sum_{j=1}^{J_i} f(Z_{ij})$ is the sum of risk from all episodes on the linear predictor scale. To build up intuition,  when $f(z)=z\gamma$ for a scalar $\gamma$, the accumulated risk is $\sum_{j=1}^{J_i} f(Z_{ij})=(\sum_{j=1}^{J_i} Z_{ij})\gamma$, where $T_i=\sum_{j=1}^{J_i} Z_{ij}$ is the total exposure time for patient $i$. This means that using the total duration of exposure as a covariate in an ordinary generalized linear model (GLM) is a special case of FLAME when the RAF is assumed to be linear. 

FLAME allows for different individuals with the same total duration of exposure to have different risk depending on the accumulation pattern. For example, for an individual with $60$ episodes of one-minute exposure, the risk contribution is $\sum_{j=1}^{J_i=60} f(1)=60\times f(1)$. For another individual with $1$ episode of sixty-minute exposure, the risk contribution is $\sum_{j=1}^{J_i=1} f(60)=f(60)$. If $f(60) > 60\times f(1)$, we conclude that a sixty-minute episode is riskier than $60$ one-minute episodes. FLAME also allows us to estimate $f(60) - 60\times  f(1)$, $f(60)/[60\times f(1)]$, and  $[f(60)-60\times f(1)]/f(1)$, all of which are measures of how much riskier a sixty-minute episode is compared to $60$ one-minute episodes. Therefore, model~\eqref{eqn:FLAME} is designed to answer the specific scientific question described in Section \ref{sec:intro}. The RAF $f(\cdot)$ is modeled nonparametrically using penalized splines to provide greater flexibility in capturing complex shapes.

\subsection{Relation to Other Methods}\label{sec:relation}
Standard generalized additive model (GAM) formulations typically operate on a scalar summary of exposure, such as total duration, and are therefore not designed to model risk accumulation at the level of individual exposure episodes.

Functional data approaches treat exposure as a continuous trajectory over time and model its effect via time-indexed coefficient functions. However, standard functional data methods assume a common, fixed domain across subjects, an assumption that is violated in our setting because surgery duration varies across patients. Extensions to uneven functional domains \citep{gellar2014,Johns02102019} can be applied to raw MAP measurements or dichotomized hypotension indicators, but their goal is to identify the optimal way of weighting time-indexed measurements to obtain the total contribution of MAP towards the response.
As such, they do not address duration-dependent risk accumulation at the episode level. Consequently, these approaches are not naturally aligned with settings where exposure occurs in irregular, discrete episodes, for which \emph{duration} -- rather than clock time -- is the primary driver of risk.

Distributed lag models (DLMs) aim to assess delayed and temporally weighted total effects of time-varying exposure \emph{intensity} by explicitly modeling associations over a prespecified lag window. Let $u_i(t)$ denote the observed exposure intensity for subject $i$ at time $t \in \mathcal{T}_i = \{1,2,\ldots,T_i\}$. The distributed lag linear model can be written as $g\{\E[y_i]\} = \beta\int_{l_0}^L w(l)u_i(t-l)dl$, where $w(l)$ denotes lag weights, $l_0$ and $L$ define the minimum and maximum lags considered, and $\beta$ captures the association between the weighted cumulative exposure and the outcome. Distributed lag non-linear models (DLNMs) extend this model to accommodate non-linear exposure--response relationships. One class of DLNMs uses a bivariate exposure--lag--response function $f\cdot w(\cdot, \cdot)$ to jointly model non-linear effects across exposure intensity and lags as $\int_{l_0}^L f\cdot w\big(u_i(t-l), l\big) dl$ \citep{armstrong2006models, gasparrini2010distributed, gasparrini2014modeling, gasparrini2017penalized}, while another applies a non-linear mapping $f(\cdot)$ to the lag-weighted cumulative exposure intensity so that the effect is modeled as $f\!\left\{\int_{l_0}^L w(l) u_i(t-l)dl \right\}$ \citep{wilson2022kernel, wang2023semiparametric, pan2025estimating}. \citet{pan2025estimating} refer to the first type as  distributed response function DLNMs (DRF-DLNMs) and the second type as adaptive cumulative exposure DLNMs (ACE-DLNMs).     

Despite their flexibility, several key distinctions separate DLMs/DLNMs from FLAME. First, DLMs and DLNMs focus on estimating the effect of either lag-weighted cumulative exposure intensity or lag-specific exposure intensity, whereas FLAME is motivated by settings in which exposure histories consist of irregular, discrete episodes with variable \emph{durations}. The primary scientific question in FLAME concerns how risk accumulates as a function of episode duration, rather than how exposure intensity propagates over time through lagged effects. Consequently, FLAME targets duration-dependent accumulation mechanisms that are not naturally represented within the DLM/DLNM framework.

One might argue that episode duration could be treated as  exposure intensity so that DLM/DLNM can be applied. However, this leads to ambiguity in defining lag. In DLM/DLNM applications, lag is defined relative to a well-defined time unit (e.g., daily exposure to pollutants), making interpretation straightforward. In contrast, when duration is treated as intensity, lag could be referenced to the beginning, the end, or any point within an exposure episode. This ambiguity complicates both model specification and interpretation.

A second distinction is that DLMs/DLNMs assume that all subjects share a common temporal grid and thus the same number of opportunities to experience exposure. In contrast, FLAME does not have such restrictions and can accommodate subject-specific exposure histories with varying numbers of episodes. In our application, surgery duration differs across patients. A patient who had a six-hour surgery, for example, does not have the opportunity to experience any hypotension event beyond six hours. Padding shorter trajectories with zeros in the DLM/DLNM framework is problematic, as each entry on the lag grid is interpreted as an observed exposure value at a specific lag. As a result, padding unobserved portions of the lag grid with zeros conflates ``zero exposure intensity" with ``no opportunity for exposure". Such conflation can lead to biased estimation of the exposure–lag–response relationship.

A third distinction is that DLMs/DLNMs assume that the timing of disease onset is both known and of scientific interest, so that lag can be meaningfully defined and interpreted. In contrast, FLAME does not model lagged effects and therefore does not rely on this requirement. In our application, we are interested in the standard clinical outcome, ``AKI within 48 hours", defined as a binary indicator of whether a patient develops AKI within 48 hours after surgery. The exact time of AKI onset may vary substantially across patients, occurring 1 hour or 48 hours after surgery, but this timing is neither observed nor of primary interest for the AKI analysis. As a result, defining lag as the time from the beginning of a hypotensive episode to a fixed reference point (e.g., 48 hours post-surgery) is difficult to interpret. When lag itself lacks a clear clinical or scientific meaning, the resulting exposure–lag–response relationship also becomes challenging to interpret, limiting the utility of DLMs/DLNMs in our setting.

In summary, FLAME is designed for settings with irregular, episodic exposures, where the primary scientific question concerns how risk accumulates as a function of episode duration, rather than exposure intensity or lagged effects over time. This focus distinguishes it from existing approaches such as GAMs, functional data methods, and DLMs/DLNMs, which are not naturally suited to addressing this question. 

\subsection{Parameter Estimation and Inference}
\label{sec:esti}
Model \eqref{eqn:FLAME} can be efficiently fitted within the generalized additive modeling framework using the \texttt{mgcv} package. Following reviewers’ suggestion, we represent $\sum_{j=1}^{J_i} f(Z_{ij})$ as a linear functional of a smooth, which enables direct use of existing software for penalized spline estimation, inference, and model checking.

Specifically, we organize the vectors of duration of exposure episodes $\boldsymbol{Z}_1, \boldsymbol{Z}_2, ..., \boldsymbol{Z}_I$ into a matrix $\boldsymbol{Z}$ of dimension $I \times J$, where $J = \max\{J_1, J_2, ..., J_I\}$. For subjects with $J_i < J$, the remaining entries are padded with zeros, i.e., $Z_{ij} = 0$ for $j > J_i$. Under the point constraint $f(0) = 0$, applying a linear functional of the smooth to each row of $\boldsymbol{Z}$ yields the desired contribution $\sum_{j=1}^{J_i} f(Z_{ij})$ for $i = 1, ..., I$. Zero-padding is valid in FLAME because padded entries correspond to the absence of an exposure episode, and the constraint $f(0) = 0$ ensures that these entries contribute nothing to the linear predictor. 

The constraint $f(0) = 0$ is both technically convenient and substantively motivated by practice, as a zero-duration exposure should contribute no risk. Additional shape constraints may be imposed based on domain knowledge. In our application, we enforce a monotone increasing constraint, reflecting the physiological expectation that longer durations of hypotension are associated with higher risk. Other constraints, such as convexity, are well studied \citep{pya2015shape} and can be incorporated as needed.

Statistical inference for the RAF can be conducted by working with the prediction matrix $\Zbf_{pred}$ \citep{wood2017gam}. For completeness, we briefly summarize the main results. Let $\widehat\gammabf$ denote the estimated spline coefficients and $\widehat \Vbf_{\gammabf}$ their covariance matrix. The predicted linear predictor is given by $\widehat \etabf_{pred} = \Zbf_{pred} \widehat \gammabf$, with covariance matrix $\widehat \Vbf_{\etabf_{pred}} = \Zbf_{pred} \widehat \Vbf_{\gammabf} \Zbf_{pred}^T$. To compare linear predictors at two settings, let $\Zbf_{pred}$ denote the corresponding prediction matrix and $\dbf = (1, -1)^T$, then the estimated difference is $\delta = \dbf^T \etabf_{pred}$, with variance $\widehat \Var(\delta) = \dbf^T \widehat \Vbf_{\etabf_{pred}} \dbf$. Inference on the response scale can be obtained via posterior simulation \citep{wood2017gam}. Specifically, posterior samples of $\gammabf$ can be simulated from a multivariate normal distribution with mean $\widehat \gammabf$ and covariance $\widehat \Vbf_{\gammabf}$.

To summarize, FLAME offers three main advantages. First, it directly targets the scientific question of interest by modeling risk accumulation at the level of individual exposure episodes, allowing subjects with the same total exposure duration to have different risk depending on how that exposure is distributed across episodes. Second, FLAME represents risk accumulation through an unknown risk accumulation function (RAF) without imposing a specific parametric form. This nonparametric formulation yields an interpretable characterization of how risk depends on episode duration, while reducing the risk of model misspecification and allowing the data to capture complex nonlinear patterns \citep{ruppert2003semiparametric, HarezlakRuppertWand2018Book}. Third, FLAME is readily extensible to more complex scientific settings. For example, while we assume a common RAF across the entire observation window for simplicity, different phases of cardiac surgery (e.g., before, during, or after cardiopulmonary bypass) may follow distinct accumulation patterns. The framework can naturally accommodate phase-specific RAFs, an extension we plan to explore in future work. 

Although FLAME can be expressed within the generalized additive modeling framework from a computational perspective, its novelty lies not in the computational machinery, but in the problem formulation and the episode-level representation of duration-dependent risk accumulation, which are not standard in existing applications of GAMs.

The proposed method is implemented in the \texttt{R} package \texttt{flameRisk}, which supports a range of generalized outcome distributions, including Gaussian, binary, and count responses.

\section{Simulation study}\label{sec:simulation}
We conducted simulation studies to evaluate the performance of FLAME under a range of settings. Binary outcomes were generated from a model with linear predictor ${\rm logit}(p_i)=\Xbf_i\betabf + \sum_{j=1}^{J_i} f(z_{ij})$, where $\Xbf_i\betabf=-3.5+0.1X_{i1}$ and $X_{i1}\sim N(0,1)$.

The number of episodes $J_i$ was generated independently with equal probabilities over $\{0,1,\ldots,15\}$, and episode duration $z_{ij}$ were generated independently from a uniform distribution on $(0,30]$.  We varied the sample size $I \in \{1000, 2000, 5000, 10000\}$ and the dimension of basis $K \in\{30, 40, 50\}$. Four shapes were considered for the RAF $f(\cdot)$: linear, piecewise linear, logarithm, and sigmoid. Parameters of the RAF such as the slope of the linear RAF were calibrated to achieve event rates of $10\%, 30\%, \text{ and }50\%$, which constitute an additional simulation factor. The exact functional forms are provided in the Supplementary Materials.

In total, we considered $144$ simulation scenarios, each replicated $1000$ times. The estimated RAF, $\widehat{f}(z)$, was evaluated against the true function $f(z)$ using the integrated squared error (ISE), $\int_0^{30}\{\widehat f(z) - f(z)\}^2dz$. Pointwise coverage probabilities were averaged over the domain to obtain overall coverage.

\subsection{Simulation Results}\label{subsec:}
\begin{figure}
\centering
\includegraphics[width=1\textwidth]{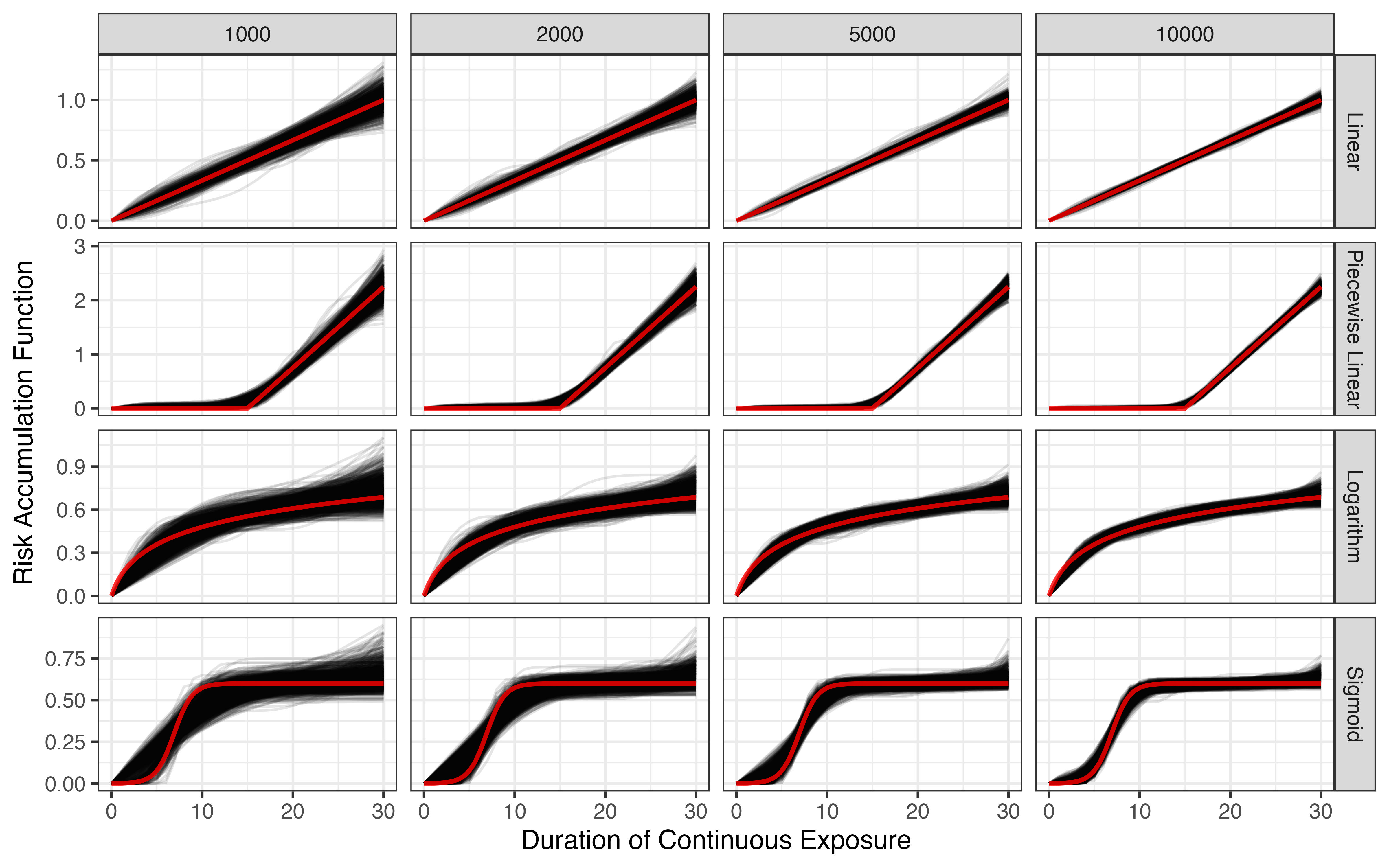}
\caption{True RAF (red) and estimated RAFs (black) based on $1{,}000$ simulated datasets for sample sizes $I = 1000, 2000, 5000, 10000$ under four RAF shapes (linear, piecewise linear, logarithmic, and sigmoid). The event rate was fixed at $50\%$, and the the dimension of basis $K = 30$.}
\label{fig:sim-raf-50}
\end{figure}

Figure~\ref{fig:sim-raf-50} displays the estimated RAFs across different sample sizes for linear, piecewise linear, logarithm, and sigmoid shapes. Results are shown for the event rate of $50\%$ and basis dimension $K=30$. In each panel, the red curve represents the true function, 
while the black curves correspond to estimates from simulated datasets. As expected, estimation accuracy improves with increasing sample size across all settings.
Notably, linear and piecewise linear RAFs are well recovered even at smaller sample sizes, whereas logarithm and sigmoid shapes require larger sample sizes to accurately capture their features.

Table \ref{tab:sim-ise-cvg-width-er50} summarizes the ISE, mean pointwise 95\% confidence interval coverage, and average interval width across $1,000$ simulation replicates. Both ISE and interval width decrease quickly as sample size increases in all settings. Coverage approaches the nominal level of 0.95 for linear, piecewise linear, and sigmoid RAFs, while convergence is slower for the logarithmic case. Further investigation (Supplementary Materials Section S2.1) indicates that the slower convergence in coverage for the logarithmic RAF is mainly driven by a slight downward bias at shorter durations. This bias diminishes as sample size increases, although at a slower rate. Additional simulation results, including simulations at event rates of $10\%$ and $30\%$, varying basis dimensions $K$, and computational time, are provided in the Supplementary Materials.

\begin{table}
\centering
\begin{adjustbox}{width=\textwidth}
\begin{tabular}[t]{rcccccccccccc}
\toprule
 & \multicolumn{3}{c}{Linear} & \multicolumn{3}{c}{Piecewise Linear} & \multicolumn{3}{c}{Logarithm} & \multicolumn{3}{c}{Sigmoid}\\
 \cmidrule(lr){2-4}\cmidrule(lr){5-7}\cmidrule(lr){8-10}\cmidrule(lr){11-13}
I & ISE & Cvg & Width & ISE & Cvg & Width & ISE & Cvg & Width & ISE & Cvg & Width\\
\midrule
1,000 & 0.053 & 0.941 & 0.138 & 0.221 & 0.930 & 0.332 & 0.125 & 0.844 & 0.219 & 0.184 & 0.716 & 0.216\\
\hline
2,000 & 0.026 & 0.943 & 0.098 & 0.122 & 0.937 & 0.256 & 0.071 & 0.868 & 0.172 & 0.104 & 0.802 & 0.187\\
\hline
5,000 & 0.011 & 0.942 & 0.063 & 0.055 & 0.942 & 0.181 & 0.037 & 0.881 & 0.124 & 0.036 & 0.902 & 0.149\\
\hline
10,000 & 0.005 & 0.944 & 0.046 & 0.031 & 0.942 & 0.139 & 0.022 & 0.887 & 0.097 & 0.017 & 0.935 & 0.120\\
\bottomrule
\end{tabular}
\end{adjustbox}
\caption{Simulation results across sample sizes $I$ and RAF shapes. Reported metrics include the integrated squared error (ISE), mean pointwise 95\% confidence interval coverage (Cvg), and average interval width (Width), based on $1{,}000$ simulation replicates. The event rate was fixed at $50\%$ and the dimension of basis $K = 30$.}
\label{tab:sim-ise-cvg-width-er50}
\end{table}

\section{Application}\label{sec:real-data}

\subsection{Data description}\label{subsec:data_description}
We applied FLAME to data from patients who underwent CABG surgery at the Johns Hopkins Hospital (JHH) between July 1, 2016 and October 31, 2019. Patients were included if they underwent isolated CABG with cardiopulmonary bypass (CPB). High-frequency hemodynamic measurements of MAP were extracted from the electronic health record and linked to clinical variables from the JHH Society of Thoracic Surgery (STS) database. The dataset contains $1{,}199$ patients and has been described in detail by \citet{goeddel2024mapcvp}. The outcome of interest is acute kidney injury (AKI) within 48 hours after surgery as defined by the KDIGO criteria using change in serum creatinine \citep{khwaja2012kdigo}. 

From this initial dataset, three patients with missing STS renal failure scores, a composite preoperative risk measure incorporating a wide range of demographic and clinical risk factors,  were excluded, resulting in an analytic sample of $1{,}196$ patients. Among these, $335$ ($28\%$) developed AKI within 48 hours after surgery. The exposure is hypotension, a hemodynamic state defined as MAP less than $65$ mm Hg. MAP is the average blood pressure  during a single cardiac cycle and is measured through catheters placed in the radial/femoral artery, recorded at 1-minute resolution during surgery. Seventeen patients had missing MAP values (mean (SD)  duration of missingness: 34 (20) minutes). In the primary analysis, missing values were imputed using linear interpolation; a sensitivity analysis excluding these patients is reported in the Supplementary Materials Section S3.2. We did not consider measurement error modeling for MAP as arterial MAP is widely regarded as the clinical reference standard in perioperative monitoring.

Previous studies have shown that total duration of hypotension during surgery is associated with AKI \citep{Scott2024perioperative}. It remains unclear, however, how the number and duration of hypotensive episodes influence AKI risk. For example, is a single sustained hypotensive episode more harmful than multiple shorter episodes with the same total duration? In our data, we identified a total of $26{,}044$ hypotensive episodes across all patients. Summary statistics for episode frequency and duration, along with other patient characteristics, are provided in Table~\ref{tab:data-summary}.

\begin{table}
\centering
\begin{tabular}[t]{lcc}
\toprule
\multirow{2}{*}{}  & AKI No & AKI Yes\\
                      & (N = 861) & (N = 335)\\
\midrule
Age  & 64.0 (9.6) & 65.7 (9.7)\\
Gender: Male (\%) & 676 (78.5) & 250 (74.6)\\
Renal Failure Score & 0.02 (0.04) & 0.05 (0.07)\\
Total Surgery Time (Hours) & 6.5 (1.2) & 6.9 (1.8)\\
Number of Episodes & 22 (9) & 21 (9)\\[5pt]
Number (\%) of subjects \\ with at least one episode & & \\
\;\;\;\; $\geq 10$ minutes & 620 (72.0) & 265 (79.1)\\
\;\;\;\; $\geq 20$ minutes & 261 (30.3) & 146 (43.6)\\
\;\;\;\; $\geq 30$ minutes & 105 (12.2) & 85 (25.4)\\
\;\;\;\; $\geq 40$ minutes & 48 ( 5.6) & 55 (16.4)\\
\;\;\;\; $\geq 50$ minutes & 34 ( 3.9) & 29 ( 8.7)\\
\;\;\;\; $\geq 60$ minutes & 24 ( 2.8) & 21 ( 6.3)\\
\bottomrule
\end{tabular}
\caption{Summary of the analytic sample. Gender and episode distribution  are summarized as count (\%). All others variables are summarized as mean (SD).}
\label{tab:data-summary}
\end{table}

\subsection{Analysis results}

\begin{figure}[!tbh]
\centering
\includegraphics[width=0.5\textwidth]{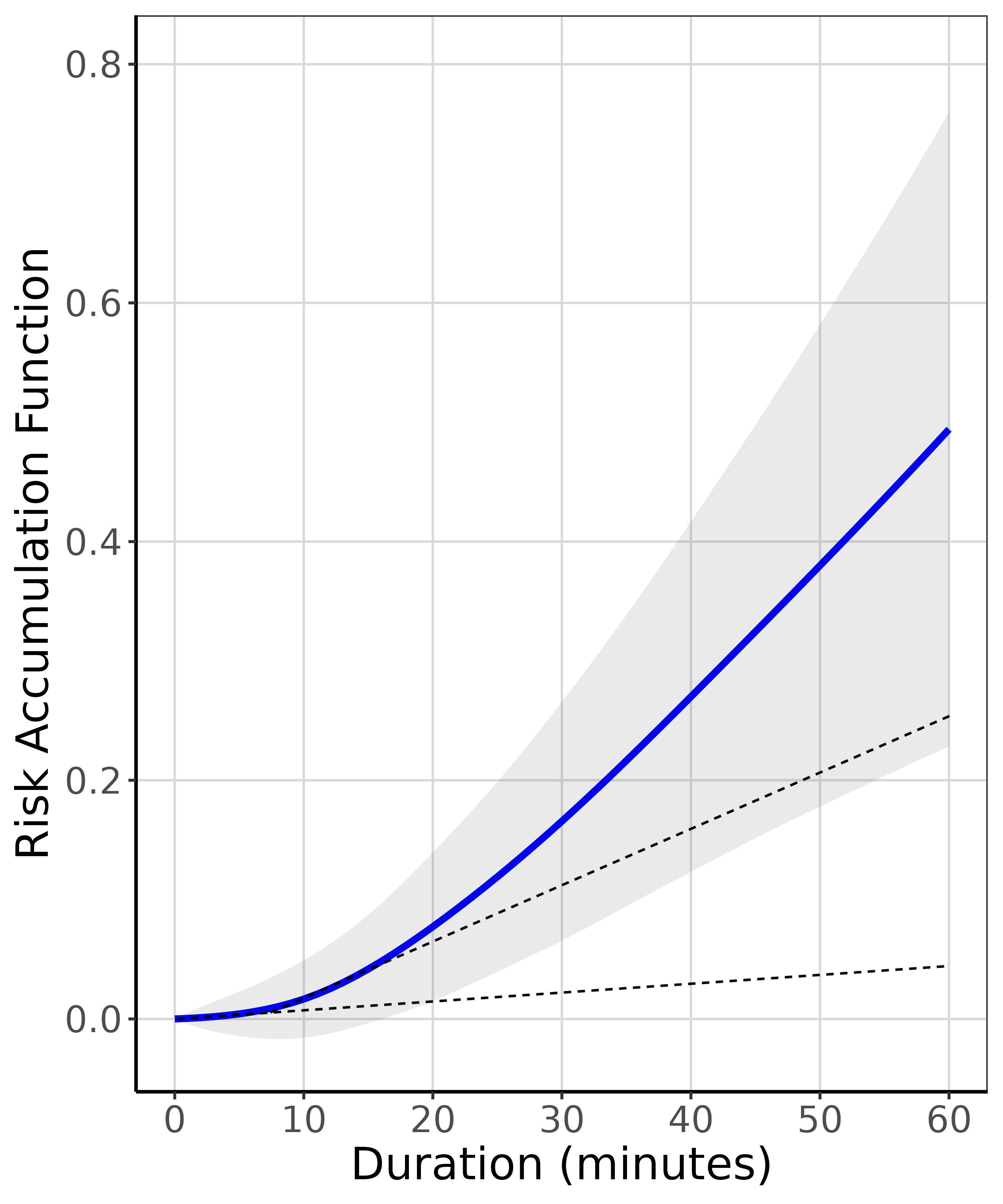}
\caption{Estimated AKI risk accumulation function (RAF) for hypotension duration using FLAME (blue line), with pointwise 95\% confidence intervals shown as the gray shaded region. The dashed lines represent linear extrapolation of the initial slope for comparison.}
\label{fig:real-data-raf}
\end{figure}

We applied FLAME to these data, adjusting for baseline covariates including total surgery duration and the STS predicted renal failure score. Figure~\ref{fig:real-data-raf} displays the estimated RAF (blue curve) with 95\% confidence intervals (gray shaded region). The dashed lines show linear extrapolation of the initial slope of the RAF for comparison. The estimated RAF suggests that risk accumulates slowly and approximately linearly for episodes shorter than 5 minutes, increases more rapidly for durations between 5 and 15 minutes, and becomes even steeper for longer episodes. This pattern indicates that a single prolonged hypotensive episode may pose greater AKI risk than multiple shorter episodes with the same total duration.

To formally assess this hypothesis, we estimated AKI probabilities under seven scenarios, each with a total hypotension duration of 60 minutes, and compared them to a reference scenario consisting of a single 60-minute episode. The results are presented in Table~\ref{table:aki-risk}. Given the exploratory nature of this analysis, no adjustment for multiple comparisons was performed.

Despite having the same total duration, scenarios with longer individual episodes generally exhibit higher estimated AKI risk. Compared to the reference scenario, three scenarios -- 60 one-minute episodes, 12 five-minute episodes, and 6 ten-minute episodes -- show statistically significant differences. We note, however, that relatively few episodes exceed 20 minutes in duration, leading to increased uncertainty in this region, as reflected by wider confidence intervals. As such, estimates for longer durations should be interpreted with caution. Larger studies are needed to confirm these findings.

To assess model adequacy and contextualize these findings, we compared FLAME with simpler alternatives based on total hypotension duration. A generalized additive model using total duration yielded a higher AIC ($1343.45$) than FLAME ($1333.92$), indicating improved fit for FLAME. In addition, models that included fragmentation measures (e.g., number of hypotension episodes and variability in episode duration) alongside total duration provided evidence that, even after controlling for total duration, episodic structure contributes to risk. These results support the importance of modeling duration-dependent accumulation beyond scalar summaries. Detailed results are provided in the Supplementary Materials Sections S3.3 and S3.4.

Additional analyses, including RAF estimates over the full duration range, a sensitivity analysis excluding patients with missing MAP values, and comparisons with alternative methods, are provided in the Supplementary Materials Section S3.

\begin{table}[ht]
\begin{center}
\begin{adjustbox}{width=0.95\textwidth}
\begin{tabular}{lccc}
\toprule
\multicolumn{1}{c}{\,} & \multicolumn{1}{c}{Estimate for each scenario} & \multicolumn{2}{c}{Comparison with $f(60)$}\\ 
\cmidrule(lr){2-2} \cmidrule(lr){3-4}
Scenario & Estimate (95\% CI)  & Difference (95\% CI) & P-value  \\ \midrule                
$60 \times f(1)$ (60 one-minute episodes) & 0.24 (0.20, 0.28) & 0.09 (0.02, 0.18) & 0.014 \\
$12 \times f(5)$ (12 five-minute episodes) & 0.24 (0.21, 0.27) & 0.09 (0.02, 0.17) & 0.007 \\
$6 \times f(10)$ (6 ten-minute episodes) & 0.25 (0.22, 0.29) & 0.08 (0.02, 0.16) & 0.012 \\
$4 \times f(15)$ (4 fifteen-minute episodes) & 0.27 (0.22, 0.31) & 0.07 (0.01, 0.14) & 0.028 \\
$3 \times f(20)$ (3 twenty-minute episodes) & 0.28 (0.23, 0.33) & 0.06 (0.00, 0.12) & 0.057 \\
$2 \times f(30)$ (2 thirty-minute episodes) & 0.30 (0.24, 0.36) & 0.04 (-0.01, 0.09) & 0.157 \\
$1 \times f(60)$ (1 sixty-minute episode)  & 0.33 (0.25, 0.42) & - & - \\\midrule   
\bottomrule
\end{tabular}
\end{adjustbox}
\end{center}
\caption{Estimated AKI probabilities for seven scenarios, along with comparisons to a reference scenario consisting of a single 60-minute episode. Covariates, including the renal failure score and total surgery duration, were fixed at their mean values.}
\label{table:aki-risk}
\end{table}

\section{Discussion}\label{sec:discussion}
We proposed FLAME, a novel semiparametric model for directly estimating duration-dependent risk accumulation at individual episode level from temporal exposures. By mapping each episode’s duration through a nonparametrically estimated RAF, FLAME avoids collapsing exposure histories into a single scalar summary such as total duration and instead captures how risk evolves as a function of exposure duration. This formulation is particularly well suited for settings where exposure occurs in irregular, discrete episodes and where prolonged continuous exposure may carry greater risk than fragmented exposure patterns. In this sense, FLAME provides a conceptually distinct alternative to existing approaches that focus on exposure intensity or lagged effects over time.

Simulations demonstrate that FLAME can recover the underlying RAF across a range of functional forms, event rates, sample sizes, and basis dimensions. In the real-data application to intraoperative hypotension during CABG surgery, FLAME reveals clinically meaningful nonlinear accumulation patterns, suggesting that prolonged hypotension is associated with disproportionately elevated AKI risk. These findings may help inform intraoperative management by highlighting the potential importance of sustained hypotension. More broadly, the estimated RAF provides an interpretable characterization of risk accumulation that is not readily captured by models based solely on total duration.

FLAME is an adaptive modeling framework and can be extended in several directions. For example, although we focus on duration of hypotension, FLAME can be adapted to incorporate hypotension intensity, such as the area under the hypotension threshold of 65 mm Hg, thereby capturing both duration and severity. The framework can also accommodate multiple RAFs to reflect heterogeneous risk accumulation across different stages of a clinical process, such as different phases of cardiac surgery. 

Several limitations should be noted. Accurate estimation of the RAF requires sufficient data across the range of exposure durations, and uncertainty may increase in regions with sparse observations. In addition, the approach relies on a meaningful definition of exposure episodes, which may vary across applications. Finally, while shape constraints can improve interpretability, they may introduce bias near boundaries in finite samples.

In conclusion, FLAME provides a flexible and interpretable approach for studying duration-dependent risk accumulation in settings with episodic exposures. By explicitly modeling how risk depends on exposure duration, the framework offers new insights beyond traditional approaches and has broad potential applications in clinical research, environmental health, and other domains involving high-resolution temporal data.


\setcounter{page}{1}
\setcounter{section}{0}
\setcounter{table}{0}
\setcounter{figure}{0}
\renewcommand{\thesection}{S.\arabic{section}}
\renewcommand{\thetable}{S.\arabic{table}}
\renewcommand{\theequation}{S.\arabic{equation}}
\renewcommand{\thefigure}{S.\arabic{figure}}
\section{Supplementary Materials}
\section{RAF for simulation}
Table \ref{supp-tab:exact-raf} summarizes the risk accumulation functions (RAFs) used in the simulation studies under a range of functional forms and target event rates. The coefficients were calibrated to achieve the desired marginal event rates.
\begin{table}[tbh]
\centering
\begin{adjustbox}{width=0.6\textwidth}
\begin{tabular}[t]{lcccccccccccc}
\toprule
Shape & Event Rate & RAF\\
\midrule
Linear & 10\% & 0.03 / 3 * x \\
Linear & 30\% & 0.065 / 3 * x \\
Linear & 50\% & 0.1 / 3 * x \\
Piecewise Linear & 10\% & 0.1 / 3 * (x - 15) * $\mathbf{1}_{\{x > 15\}}$ \\
Piecewise Linear & 30\% & 0.25 / 3 * (x - 15) * $\mathbf{1}_{\{x > 15\}}$\\
Piecewise Linear & 50\% & 0.45 / 3 * (x - 15) * $\mathbf{1}_{\{x > 15\}}$ \\
Logarithm & 10\% & 0.06 * log(x + 1) \\
Logarithm & 30\% & 0.12 * log(x + 1)\\
Logarithm & 50\% & 0.2 * log(x + 1)\\
Sigmoid & 10\% & 0.2 / (1 + 1000 * exp(-x))\\
Sigmoid & 30\% & 0.4 / (1 + 1000 * exp(-x))\\
Sigmoid & 50\% & 0.6 / (1 + 1000 * exp(-x))\\
\bottomrule
\end{tabular}
\end{adjustbox}
\caption{Risk accumulation functions (RAFs) used in the simulation studies.}
\label{supp-tab:exact-raf}
\end{table}

\section{Additional simulation results}
\label{supp-sec:additional-sim}

\subsection{Coverage for the logarithmic RAF}
In Table 1 of the main text, the coverage probability for the logarithmic RAF was observed to approach the nominal level relatively slowly. In this section, we investigate this behavior in greater detail. 

\begin{figure}[tbh]
\centering
\includegraphics[width=0.48\textwidth]{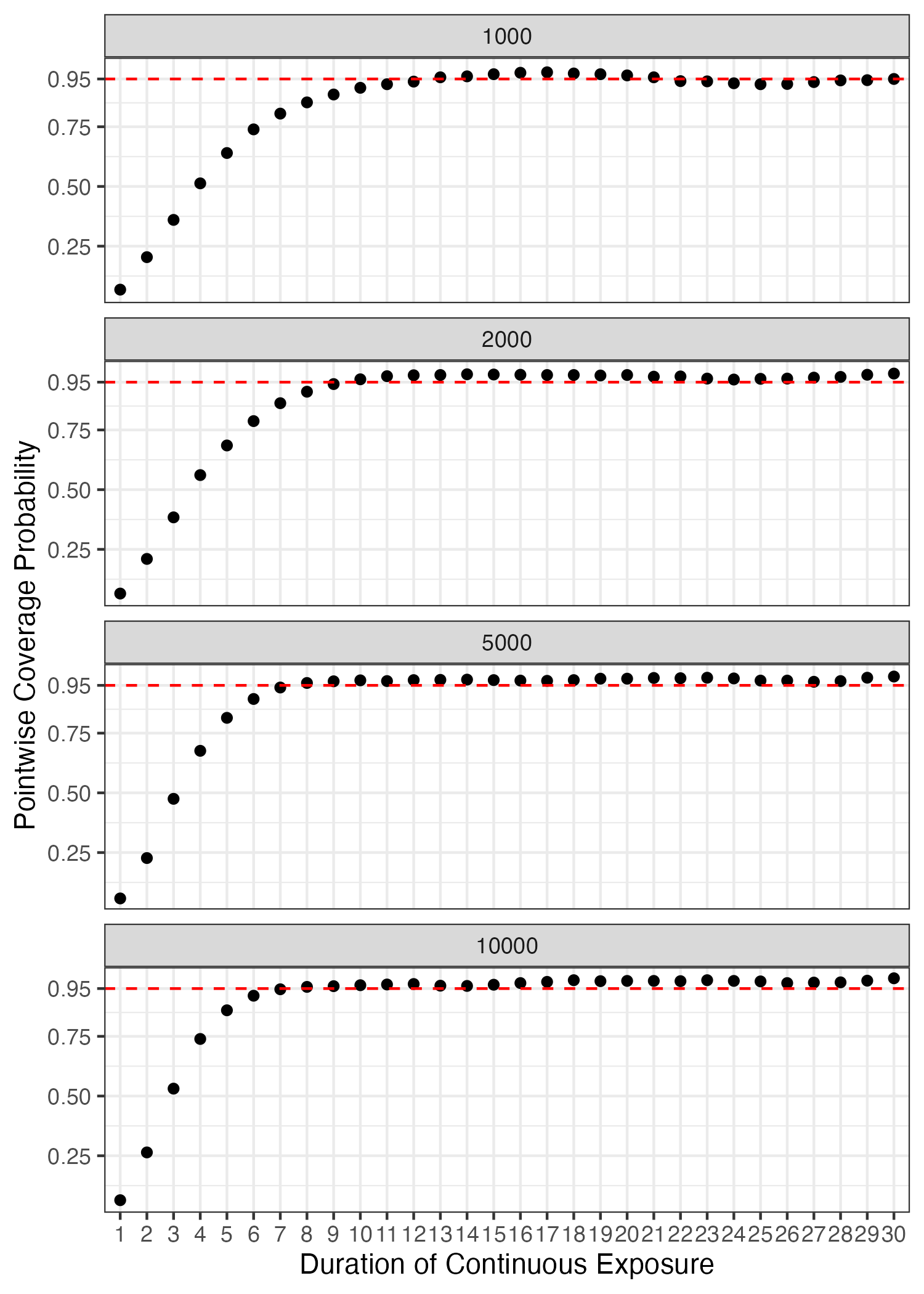}
\caption{Pointwise 95\% confidence interval coverage for the logarithmic RAF across sample sizes, with event rate fixed at 50\% and basis dimension $K=30$.}
\label{supp-fig:sim-coverage-Logarithm-er50-k30}
\end{figure}

Figure \ref{supp-fig:sim-coverage-Logarithm-er50-k30} displays the pointwise 95\% confidence interval coverage across the duration domain. Deviations from the nominal level of 0.95 are primarily observed at shorter durations: approximately $[1, 10]$ for a sample size of $1{,}000$ and $[1, 5]$ for a sample size of $10{,}000$. 

\begin{figure}
\centering
\includegraphics[width=0.45\textwidth]{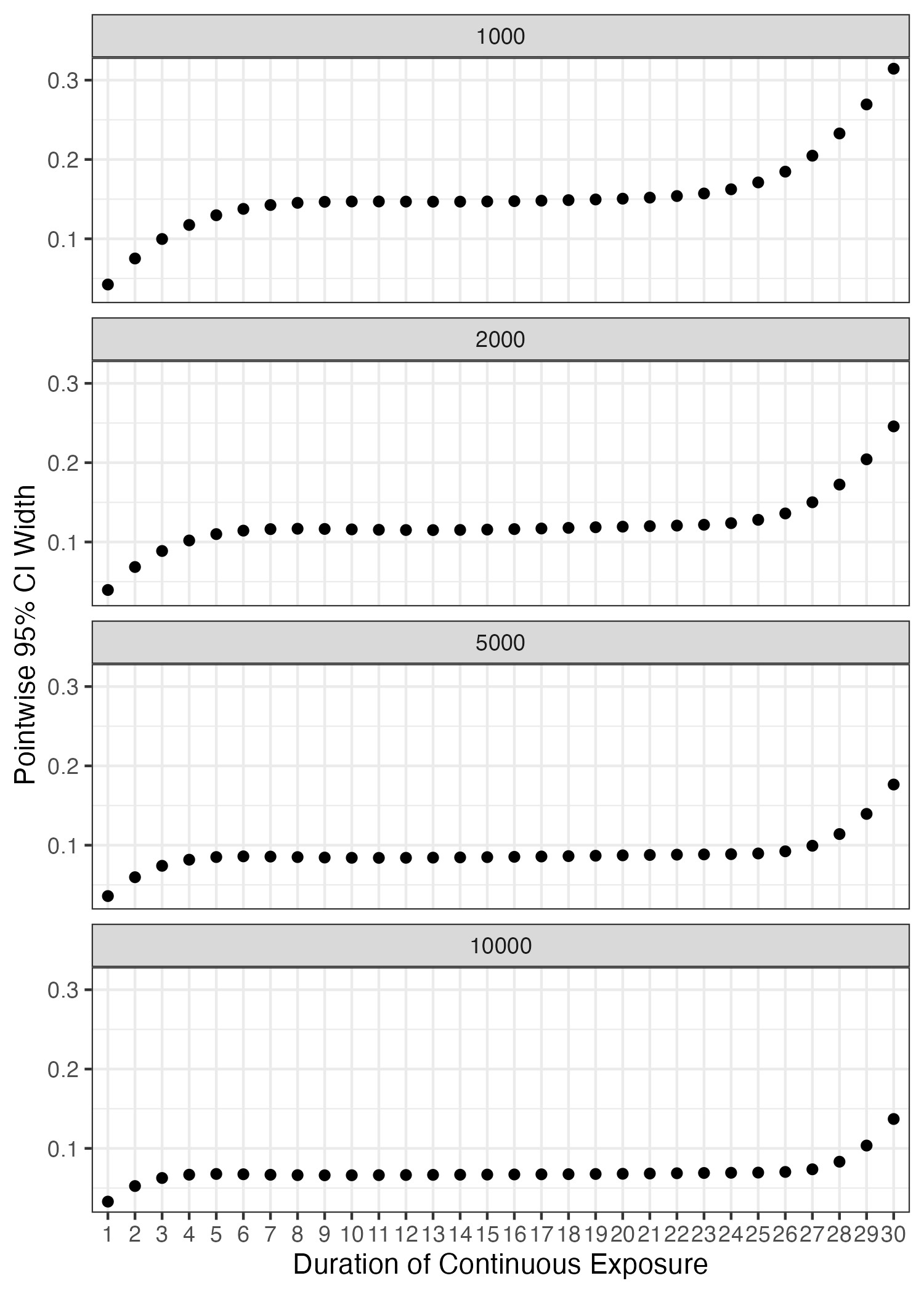}
\includegraphics[width=0.45\textwidth]{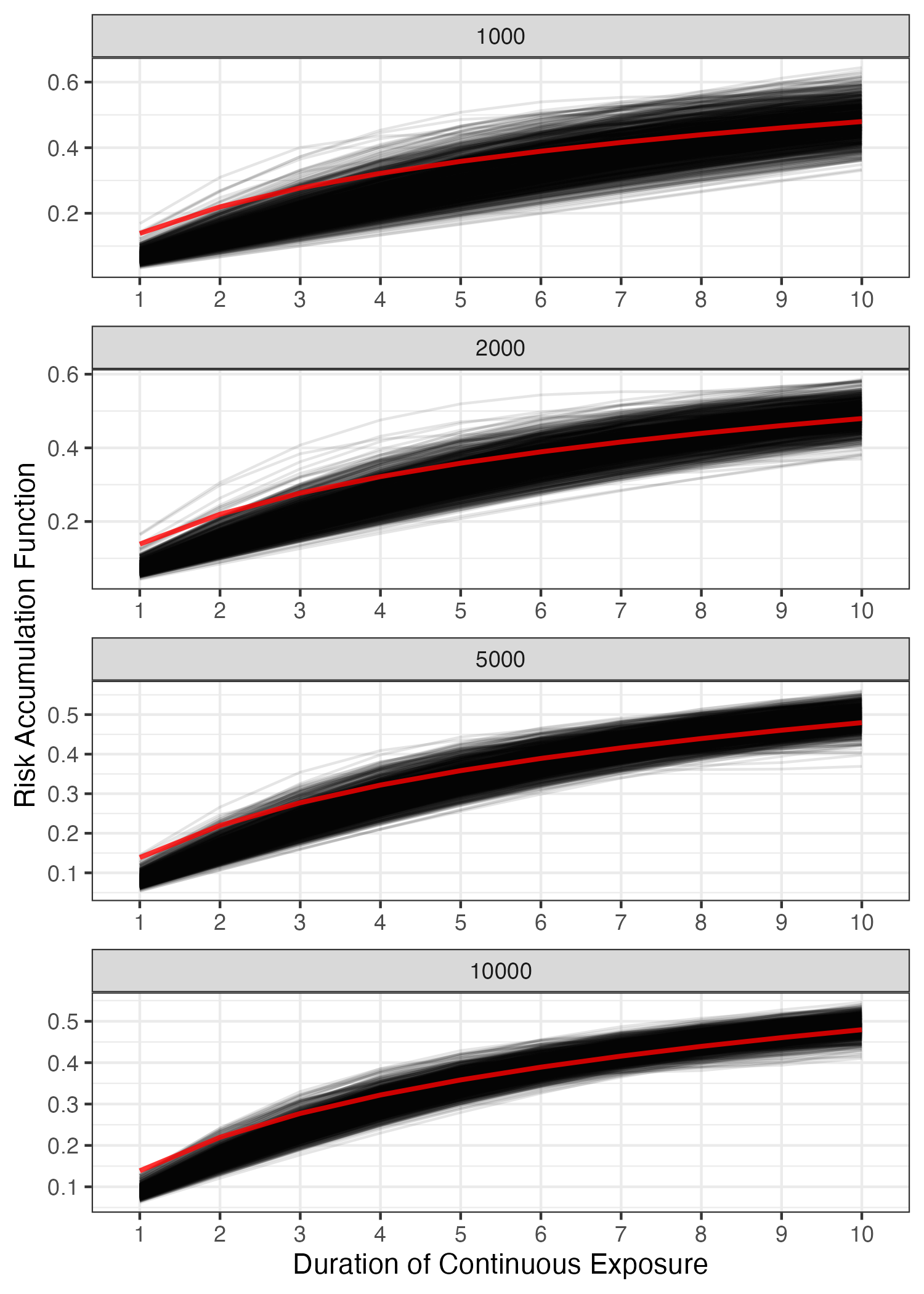}
\caption{Left: pointwise 95\% confidence interval width across sample sizes, with event rate fixed at 50\% and basis dimension $K=30$. Right: corresponding zoomed-in RAF estimates from $1,000$ simulation replicates (black) compared with the true function (red).}
\label{supp-fig:sim-CI-width-RAF-Logarithm-er50-k30}
\end{figure}

Figure \ref{supp-fig:sim-CI-width-RAF-Logarithm-er50-k30} suggests that this under-coverage is driven by a slight downward bias in the estimates, with narrower confidence intervals likely also contributing. These phenomena are likely related to the shape constraints imposed on the smooth, namely the point constraint $f(0)=0$ and the monotonicity constraint that the RAF is increasing. Both constraints are motivated by physiological considerations.

From a practical standpoint, when the estimated RAF exhibits a logarithmic shape, pointwise confidence intervals near the lower boundary should be interpreted with caution, particularly in moderate samples. When scientifically appropriate, sensitivity analyses under alternative shape constraints may help assess the robustness of inference in this region.

\subsection{Results for other event rates and basis dimensions}
Tables \ref{supp-tab:sim-ise-cvg-width-er10} and \ref{supp-tab:sim-ise-cvg-width-er30} present simulation results for event rates of $10\%$ and $30\%$. The patterns in integrated squared error (ISE), coverage, and confidence interval width are consistent with those observed for the 50\% event rate reported in the main text.

\begin{table}
\centering
\begin{adjustbox}{width=\textwidth}
\begin{tabular}[t]{crcccccccccccc}
\toprule
 & & \multicolumn{3}{c}{Linear} & \multicolumn{3}{c}{Piecewise Linear} & \multicolumn{3}{c}{Logarithm} & \multicolumn{3}{c}{Sigmoid}\\
 \cmidrule(lr){3-5}\cmidrule(lr){6-8}\cmidrule(lr){9-11}\cmidrule(lr){12-14}
K & I & ISE & Cvg & Width & ISE & Cvg & Width & ISE & Cvg & Width & ISE & Cvg & Width\\
\midrule
\multirow{4}{*}{30} 
 & 1000 & 0.032 & 0.952 & 0.111 & 0.090 & 0.866 & 0.175 & 0.054 & 0.730 & 0.123 & 0.071 & 0.667 & 0.126\\
 & 2000 & 0.016 & 0.955 & 0.080 & 0.047 & 0.908 & 0.142 & 0.033 & 0.755 & 0.101 & 0.050 & 0.651 & 0.102\\
 & 5000 & 0.007 & 0.958 & 0.052 & 0.021 & 0.924 & 0.104 & 0.019 & 0.794 & 0.077 & 0.029 & 0.689 & 0.081\\
 & 10000 & 0.004 & 0.945 & 0.039 & 0.011 & 0.934 & 0.082 & 0.010 & 0.839 & 0.063 & 0.019 & 0.717 & 0.067\\
\hline
\multirow{4}{*}{40} 
 & 1000 & 0.032 & 0.949 & 0.112 & 0.096 & 0.858 & 0.175 & 0.054 & 0.732 & 0.125 & 0.071 & 0.667 & 0.126\\
 & 2000 & 0.017 & 0.947 & 0.081 & 0.046 & 0.909 & 0.141 & 0.034 & 0.754 & 0.101 & 0.047 & 0.668 & 0.104\\
 & 5000 & 0.008 & 0.951 & 0.054 & 0.022 & 0.923 & 0.103 & 0.017 & 0.808 & 0.078 & 0.029 & 0.691 & 0.081\\
 & 10000 & 0.004 & 0.954 & 0.040 & 0.012 & 0.934 & 0.082 & 0.011 & 0.836 & 0.062 & 0.019 & 0.724 & 0.068\\
\hline
\multirow{4}{*}{50} 
 & 1000 & 0.033 & 0.944 & 0.113 & 0.093 & 0.857 & 0.173 & 0.052 & 0.751 & 0.128 & 0.072 & 0.669 & 0.126\\
 & 2000 & 0.017 & 0.948 & 0.083 & 0.050 & 0.901 & 0.141 & 0.035 & 0.754 & 0.102 & 0.047 & 0.668 & 0.103\\
 & 5000 & 0.007 & 0.953 & 0.055 & 0.021 & 0.923 & 0.104 & 0.017 & 0.807 & 0.078 & 0.028 & 0.696 & 0.082\\
 & 10000 & 0.004 & 0.944 & 0.041 & 0.012 & 0.931 & 0.082 & 0.010 & 0.835 & 0.063 & 0.019 & 0.728 & 0.068\\
\bottomrule
\end{tabular}
\end{adjustbox}
\caption{Simulation results across sample sizes $I$, basis dimensions $K$, and RAF shapes for the event rate of 10\%. Reported metrics include the integrated squared error (ISE), mean pointwise 95\% confidence interval coverage, and average interval width.}
\label{supp-tab:sim-ise-cvg-width-er10}
\end{table}

\begin{table}
\centering
\begin{adjustbox}{width=\textwidth}
\begin{tabular}[t]{crcccccccccccc}
\toprule
 & & \multicolumn{3}{c}{Linear} & \multicolumn{3}{c}{Piecewise Linear} & \multicolumn{3}{c}{Logarithm} & \multicolumn{3}{c}{Sigmoid}\\
 \cmidrule(lr){3-5}\cmidrule(lr){6-8}\cmidrule(lr){9-11}\cmidrule(lr){12-14}
K & I & ISE & Cvg & Width & ISE & Cvg & Width & ISE & Cvg & Width & ISE & Cvg & Width\\
\midrule
\multirow{4}{*}{30} 
 & 1000 & 0.029 & 0.948 & 0.104 & 0.091 & 0.927 & 0.218 & 0.065 & 0.822 & 0.150 & 0.107 & 0.696 & 0.158\\
 & 2000 & 0.015 & 0.937 & 0.075 & 0.052 & 0.930 & 0.169 & 0.036 & 0.850 & 0.119 & 0.067 & 0.751 & 0.134\\
 & 5000 & 0.006 & 0.953 & 0.048 & 0.024 & 0.937 & 0.121 & 0.018 & 0.870 & 0.086 & 0.028 & 0.860 & 0.110\\
 & 10000 & 0.003 & 0.953 & 0.035 & 0.014 & 0.937 & 0.094 & 0.011 & 0.879 & 0.068 & 0.012 & 0.915 & 0.091\\
\hline
\multirow{4}{*}{40} 
 & 1000 & 0.031 & 0.945 & 0.105 & 0.090 & 0.927 & 0.218 & 0.063 & 0.824 & 0.150 & 0.105 & 0.697 & 0.157\\
 & 2000 & 0.015 & 0.946 & 0.076 & 0.051 & 0.934 & 0.170 & 0.036 & 0.852 & 0.119 & 0.066 & 0.746 & 0.133\\
 & 5000 & 0.006 & 0.949 & 0.050 & 0.025 & 0.936 & 0.121 & 0.018 & 0.874 & 0.086 & 0.029 & 0.851 & 0.109\\
 & 10000 & 0.003 & 0.949 & 0.037 & 0.014 & 0.941 & 0.094 & 0.011 & 0.883 & 0.068 & 0.012 & 0.915 & 0.091\\
\hline
\multirow{4}{*}{50} 
 & 1000 & 0.031 & 0.945 & 0.109 & 0.092 & 0.928 & 0.219 & 0.062 & 0.826 & 0.150 & 0.103 & 0.702 & 0.159\\
 & 2000 & 0.015 & 0.950 & 0.077 & 0.050 & 0.934 & 0.169 & 0.034 & 0.859 & 0.119 & 0.066 & 0.745 & 0.133\\
 & 5000 & 0.006 & 0.953 & 0.052 & 0.024 & 0.938 & 0.121 & 0.019 & 0.870 & 0.086 & 0.028 & 0.860 & 0.110\\
 & 10000 & 0.004 & 0.942 & 0.038 & 0.014 & 0.941 & 0.094 & 0.011 & 0.882 & 0.068 & 0.012 & 0.912 & 0.091\\
\bottomrule
\end{tabular}
\end{adjustbox}
\caption{Simulation results across sample sizes $I$, basis dimensions $K$, and RAF shapes for the event rate of 30\%. Reported metrics include the integrated squared error (ISE), mean pointwise 95\% confidence interval coverage, and average interval width.}
\label{supp-tab:sim-ise-cvg-width-er30}
\end{table}

\begin{figure}[!tbh]
\centering
\includegraphics[width=0.8\textwidth]{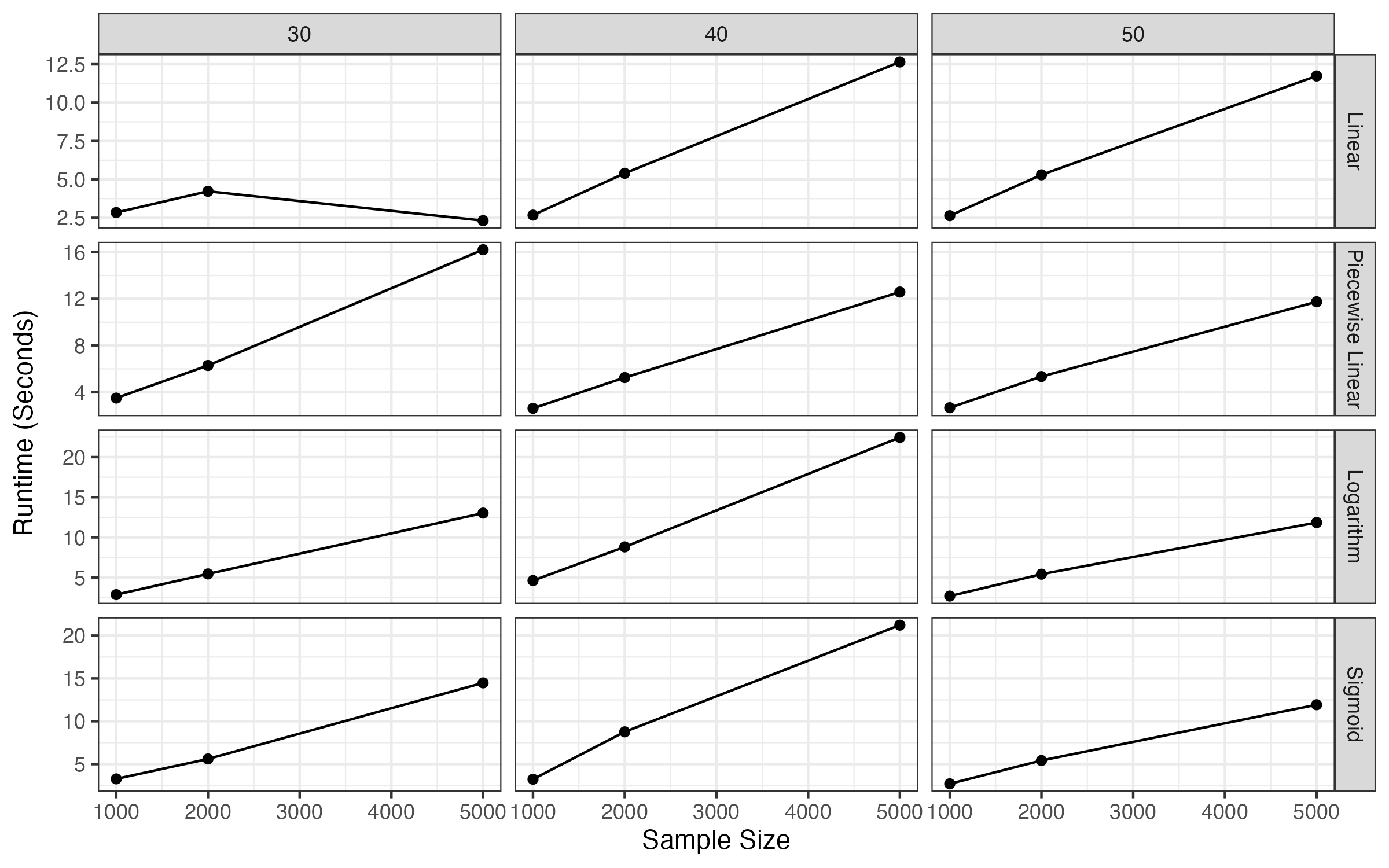}
\caption{Mean computational time across $1,000$ simulation replicates as a function of sample size, for various basis dimensions $K$ and RAF shapes, with event rate fixed at 50\%.}
\label{supp-fig:sim-runtime50}
\end{figure}

Figure \ref{supp-fig:sim-runtime50} displays the mean computational time across $1,000$ simulation replicates as a function of sample size, for various basis dimensions $K$ and RAF shapes. Computational time increases approximately linearly with sample size, demonstrating the scalability of FLAME to large datasets.

\section{Additional real data analysis results}

\subsection{Risk accumulation function for the full range}
\begin{figure}[!tbh]
\centering
\includegraphics[width=0.5\textwidth]{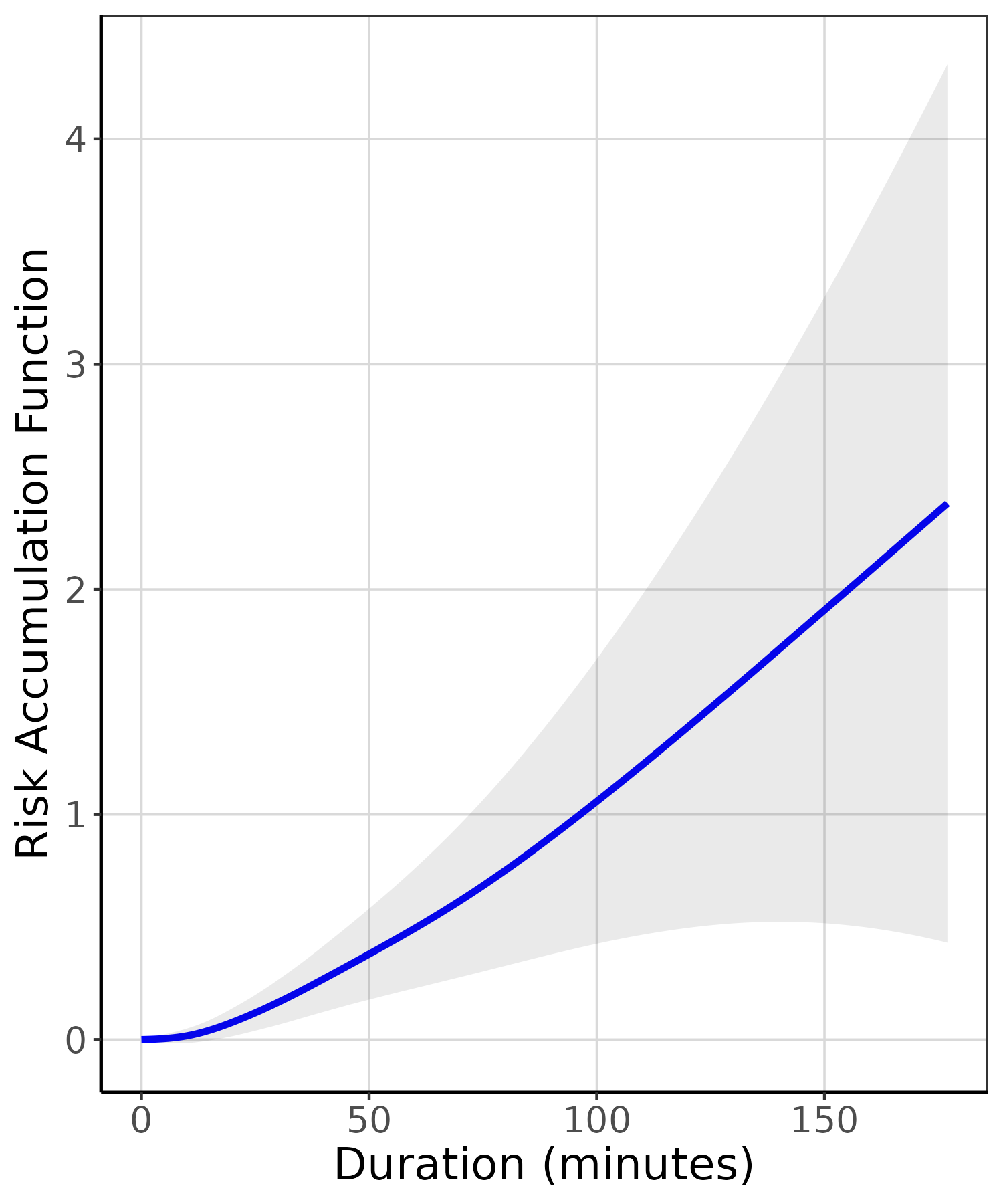}
\caption{Estimated AKI risk accumulation function (blue line) over the full range of duration values from the main analysis. The gray shaded region represents the 95\% confidence intervals.}
\label{supp-fig:real-data-raf-full-range}
\end{figure}

In the main text, we presented the AKI risk accumulation function for hypotensive episodes with durations up to 60 minutes. Here, we present the estimated RAF for the full range of observed durations in Figure \ref{supp-fig:real-data-raf-full-range}. For durations exceeding approximately 60 minutes, the RAF appears to follow a roughly linear pattern. However, limited data at longer durations lead to increased uncertainty, and additional observations in this range would be needed to improve estimation.

\subsection{Sensitivity Analysis}
\begin{figure}[!tbh]
\centering
\includegraphics[width=0.45\textwidth]{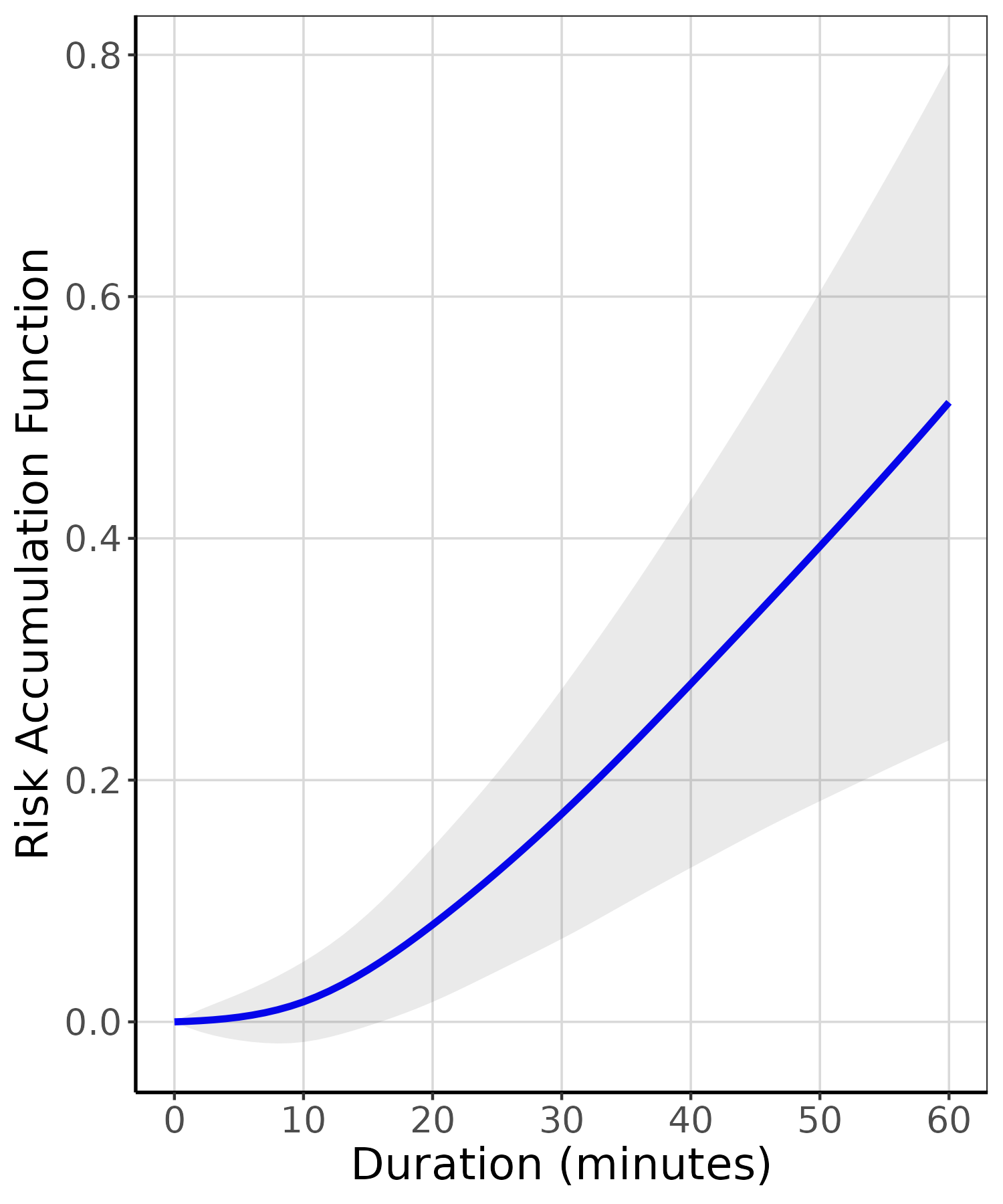}
\includegraphics[width=0.45\textwidth]{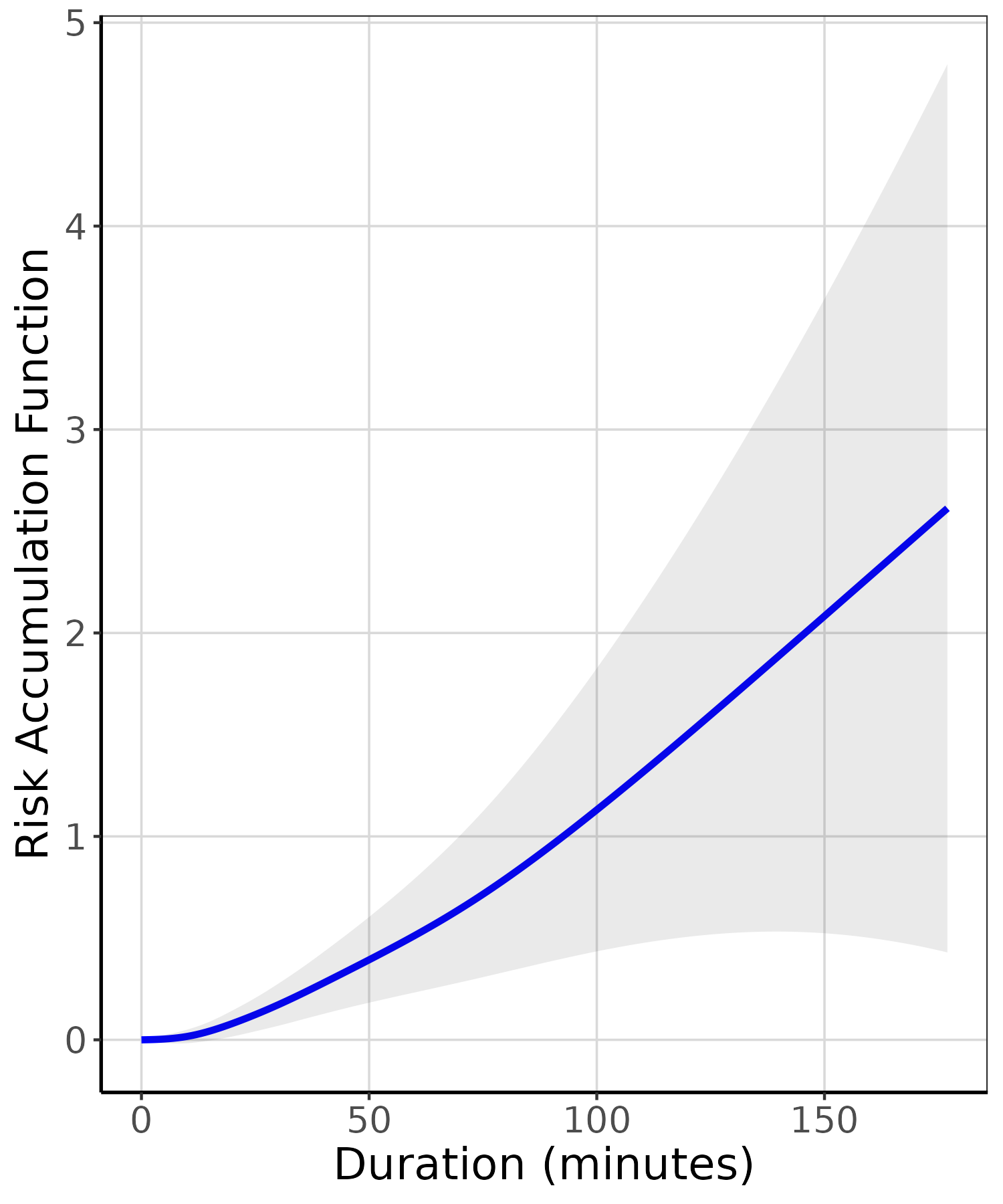}
\caption{Estimated AKI risk accumulation function (blue line) from the complete-case analysis. Left: durations up to 60 minutes. Right: full range of duration values. The gray shaded region represents the 95\% confidence intervals.}
\label{supp-fig:real-data-raf-complete-case}
\end{figure}

As a sensitivity analysis, we conducted a complete-case analysis by excluding the 17 patients with missing MAP data. Table \ref{supp-table:aki-risk-complete-case} presents results analogous to those in the main analysis. Both the estimates and statistical inferences are consistent with the main findings, suggesting that the results are robust to the interpolation strategy. Figure \ref{supp-fig:real-data-raf-complete-case} displays the estimated RAF, which closely resembles that from the main analysis.
\begin{table}[ht]
\begin{center}
\begin{adjustbox}{width=0.95\textwidth}
\begin{tabular}{lccc}
\toprule
\multicolumn{1}{c}{\,} & \multicolumn{1}{c}{Estimate for each scenario} & \multicolumn{2}{c}{Comparison with $f(60)$}\\ 
\cmidrule(lr){2-2} \cmidrule(lr){3-4}
Scenario & Estimate (95\% CI)  & Difference (95\% CI) & P-value  \\ \midrule                
$60 \times f(1)$ (60 one-minute episodes) & 0.23 (0.2, 0.27) & 0.1 (0.02, 0.19) & 0.013 \\
$12 \times f(5)$ (12 five-minute episodes) & 0.24 (0.21, 0.27) & 0.09 (0.02, 0.18) & 0.009 \\
$6 \times f(10)$ (6 ten-minute episodes) & 0.25 (0.21, 0.28) & 0.08 (0.01, 0.16) & 0.02 \\
$4 \times f(15)$ (4 fifteen-minute episodes) & 0.26 (0.22, 0.31) & 0.07 (0.00, 0.15) & 0.038 \\
$3 \times f(20)$ (3 twenty-minute episodes) & 0.28 (0.22, 0.33) & 0.06 (-0.01, 0.13) & 0.092 \\
$2 \times f(30)$ (2 thirty-minute episodes) & 0.30 (0.23, 0.36) & 0.04 (-0.02, 0.09) & 0.198 \\
$1 \times f(60)$ (1 sixty-minute episode)  & 0.33 (0.25, 0.42) & - & - \\\midrule   
\bottomrule
\end{tabular}
\end{adjustbox}
\end{center}
\caption{Estimated AKI probabilities under each scenario, with comparisons with the reference scenario of a single 60-minute hypotensive episode, based on the complete-case analysis.}
\label{supp-table:aki-risk-complete-case}
\end{table}

\subsection{Episode Fragmentation Measures}
\begin{figure}[!tbh]
\centering
\includegraphics[width=0.49\textwidth]{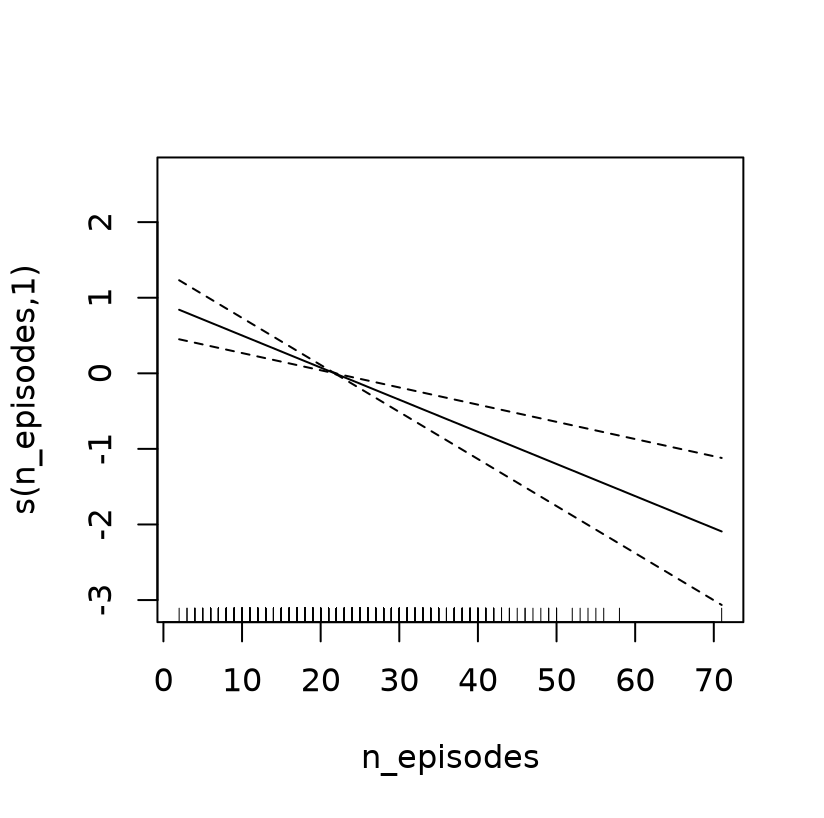}
\includegraphics[width=0.49\textwidth]{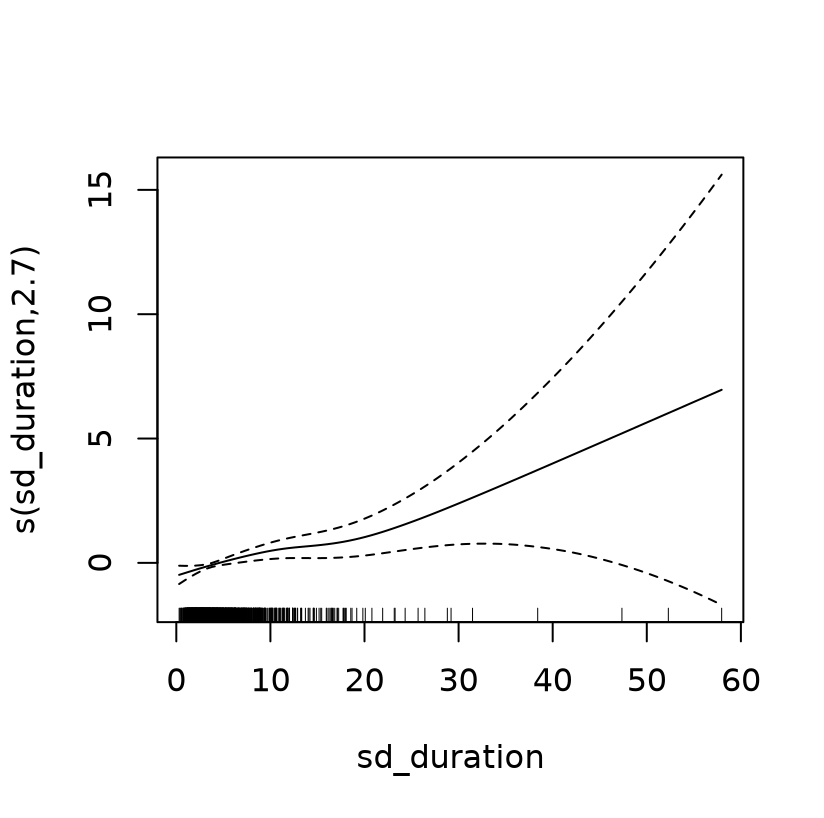}
  \caption{Evaluating the effect of fragmentation measures. Model 1 on the left shows the effect of number of hypotensive episodes; Model 2 on the right shows the effect of standard deviation of episode duration. Both Model 1 and 2 contain variables such as the predicted renal failure score, total surgery duration, and total hypotensive duration.}
  \label{supp-fig:fragmentation-measures}
\end{figure}

Motivated by the scientific question of assessing whether a long continuous episode of hypotension is more harmful to the patient than many short episodes of the same total duration, we proposed FLAME to directly model the duration-dependent risk accumulation function. To assess whether our data shows any evidence of the above, we added fragmentation measures such as the total number of episodes and the standard deviation of episode duration, and assessed these variables after controlling for the total duration of hypotension. Specifically, we fit two models, both models contain predicted renal failure score, total surgery duration, and total hypotension duration. The first model additionally contains the total number of hypotensive episodes, whereas the second model added the standard deviation of episode duration instead. Total duration of hypotension, number of hypotensive episodes, and standard deviation of episode duration are modeled using smooths, whereas other variables are kept as linear. We can see from Figure \ref{supp-fig:fragmentation-measures} (left) that holding the total hypotensive duration constant, a higher number of episodes is associated with lower AKI risk (p $< 0.001$), suggesting that more fragmented episodes are less risky than long continuous ones. Moreover, Figure \ref{supp-fig:fragmentation-measures} (right) shows that holding the total hypotensive duration constant, a higher standard deviation in the duration of episodes is associated with higher AKI risk (p = 0.002), suggesting that having a mix of episodes of both longer and shorter duration is more risky. Although neither model directly addresses the motivating scientific question, they provide empirical evidence for the importance of accounting for episode structure in addition to total duration. In this sense, they serve as simple and familiar alternatives for probing the hypothesis of interest.

\subsection{Comparison with alternative methods}
In this section, we present results from applying several alternative approaches to the real data, including generalized additive models, functional data methods, and distributed lag models.

\begin{figure}[!tbh]
\centering
\includegraphics[width=0.9\textwidth]{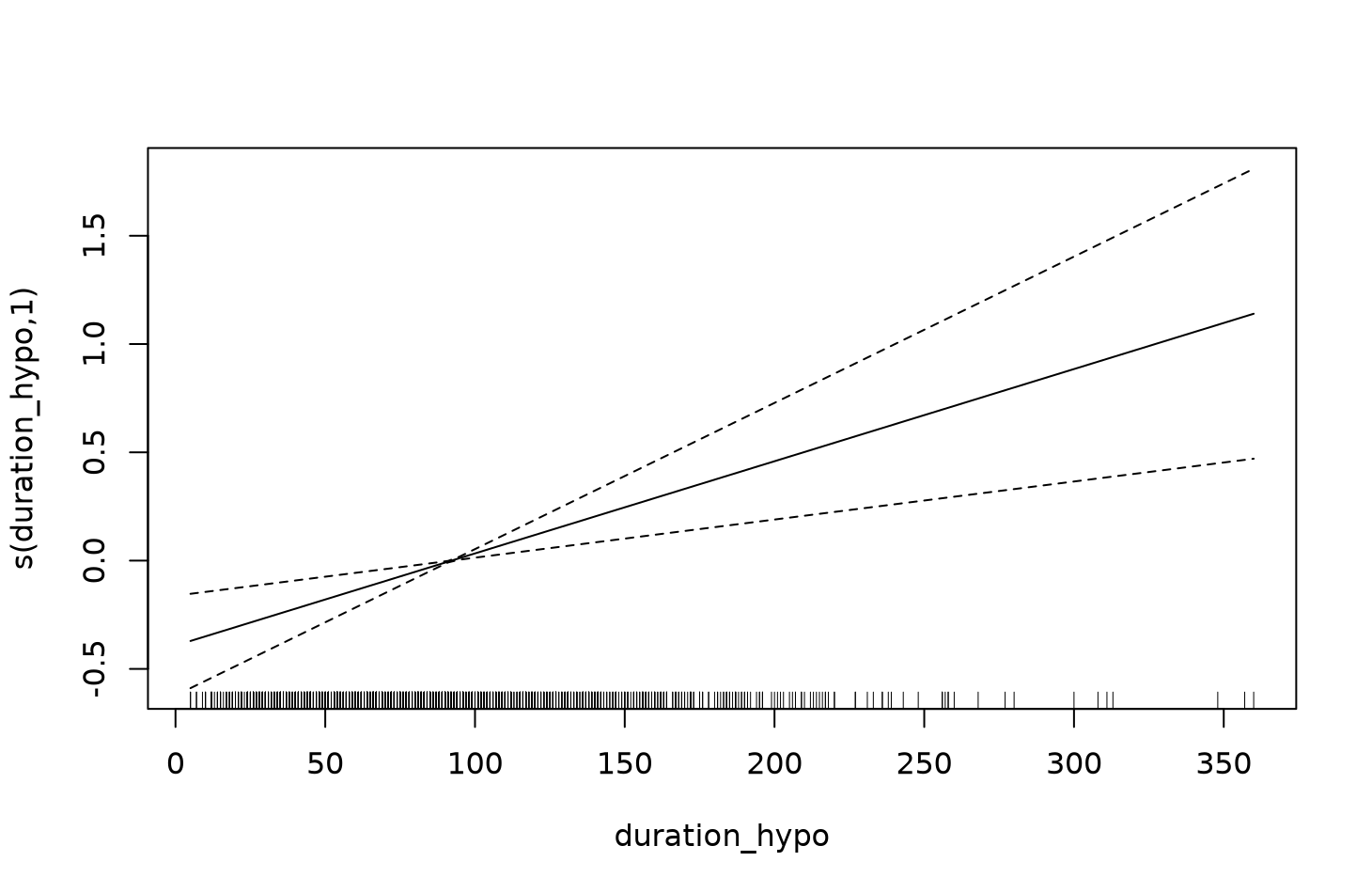}
  \caption{Estimated effect of total duration of hypotension from a generalized additive model.}
  \label{supp-fig:compare-with-gam}
\end{figure}

We begin with a generalized additive model (GAM), in which exposure is summarized as the total duration of hypotension over the course of surgery. The model was fitted using the  \texttt{gam} function from the \texttt{mgcv} package, and the resulting smooth is shown in Figure \ref{supp-fig:compare-with-gam}. While this approach captures the association between total hypotension duration and AKI risk, it does not distinguish between different episodic patterns of exposure. As a result, it does not directly address the scientific question of interest concerning duration-dependent risk accumulation at the episode level.

\begin{figure}[!tbh]
\centering
\includegraphics[width=0.9\textwidth]{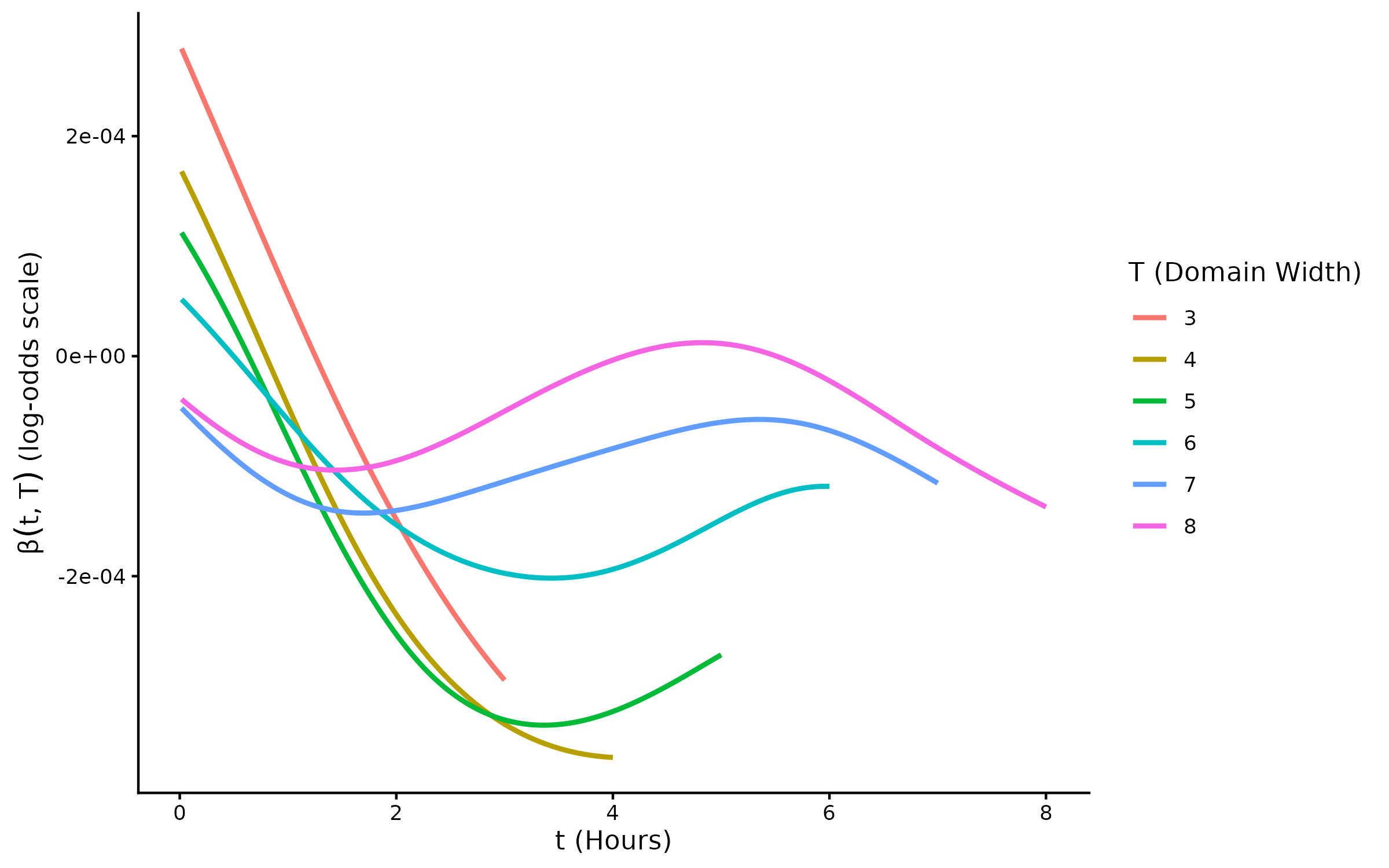}
\includegraphics[width=0.9\textwidth]{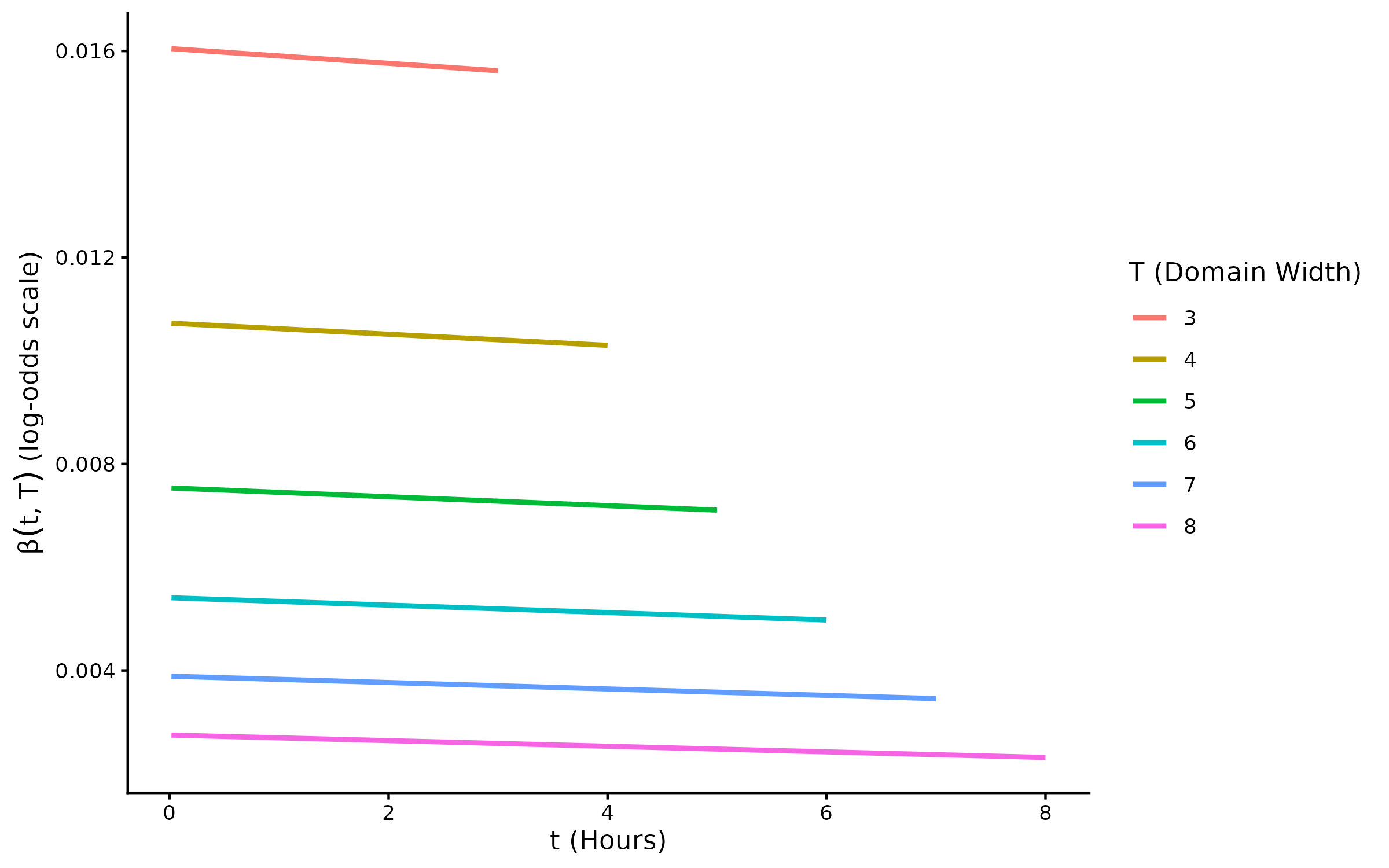}
  \caption{Model fitting results from varying domain functional regression. Top: MAP treated as a continuous variable. Bottom: MAP dichotomized as a binary indicator of hypotension (MAP $< 65$ mm Hg). }
  \label{supp-fig:compare-with-functional}
\end{figure}

Next, we tried functional data approaches. In our setting, the data can be viewed as varying-domain functional data, since surgery duration differs across patients. We therefore applied methods from \citep{gellar2014} and considered two representations of MAP: as a continuous variable and as a binary indicator of hypotension (MAP $<$ 65 mm Hg). Figure \ref{supp-fig:compare-with-functional} (top) shows the estimated coefficient functions (on the log-odds scale) for different domain widths $T$ when using the raw MAP measurements. While this approach leverages the full continuous signal, it is not specifically tailored to capture features related to hypotension. Figure \ref{supp-fig:compare-with-functional} (bottom) presents the results when MAP is dichotomized as hypotension (MAP $< 65$) at each time point. In this case, the estimated coefficient functions are approximately flat, suggesting that hypotension increases AKI risk regardless of its timing.
More broadly, functional regression methods aim to estimate a coefficient function that weights the exposure trajectory across the time domain to obtain its overall contribution to the outcome. As such, they are designed to capture time-varying effects of exposure, rather than duration-dependent accumulation at the episode level. Consequently, these approaches are not well aligned with our scientific objective.

Finally, we applied a distributed lag nonlinear model (DLNM) following the approach of \citet{gasparrini2014modeling}. To assess the ability of DLNM to address duration-dependent risk accumulation, we defined exposure as the duration of each hypotensive episode, and lag as the time from the start of each episode to the end of surgery. As discussed in Section 3.2 of the main text, this definition is not entirely natural for our setting, but is required to implement DLNM for modeling hypotension duration. In particular, ambiguity arises in defining both the start and end points of lag. For the start time, lag could be referenced to the beginning, the end, or any point within a hypotensive episode, as duration is not tied to a unique time index. For the end time, DLNM typically assumes a well-defined outcome time, but in our setting the exact timing of AKI onset is neither observed nor of primary interest.

The exposure–lag data were organized into a matrix with each row corresponding to a patient and the number of columns equal to the maximum lag across all patients, as DLNM requires a common temporal grid. Because surgery duration differs across patients, the maximum lag varies by subject, necessitating alignment to a common grid defined by the maximum lag in the sample. For time points without exposure, values were set to zero. As noted in Section 3.2, this zero-padding may introduce bias and lead to underestimated standard errors, but is necessary for model fitting within this framework.

We constructed the cross-basis using the \verb|crossbasis| function from the \verb|dlnm| package and fitted the model using the \verb|gam| function from \verb|mgcv|. We explored a range of basis specifications. However, due to the large number of zero entries in the exposure–lag matrix, spline-based bases for the exposure dimension led to unstable estimation, and we therefore adopted a linear specification for exposure dimension. For the lag dimension, we considered both cubic and penalized spline bases and observed that results were sensitive to the choice of basis and degrees of freedom.

\begin{figure}[!tbh]
\centering
\includegraphics[width=0.48\textwidth]{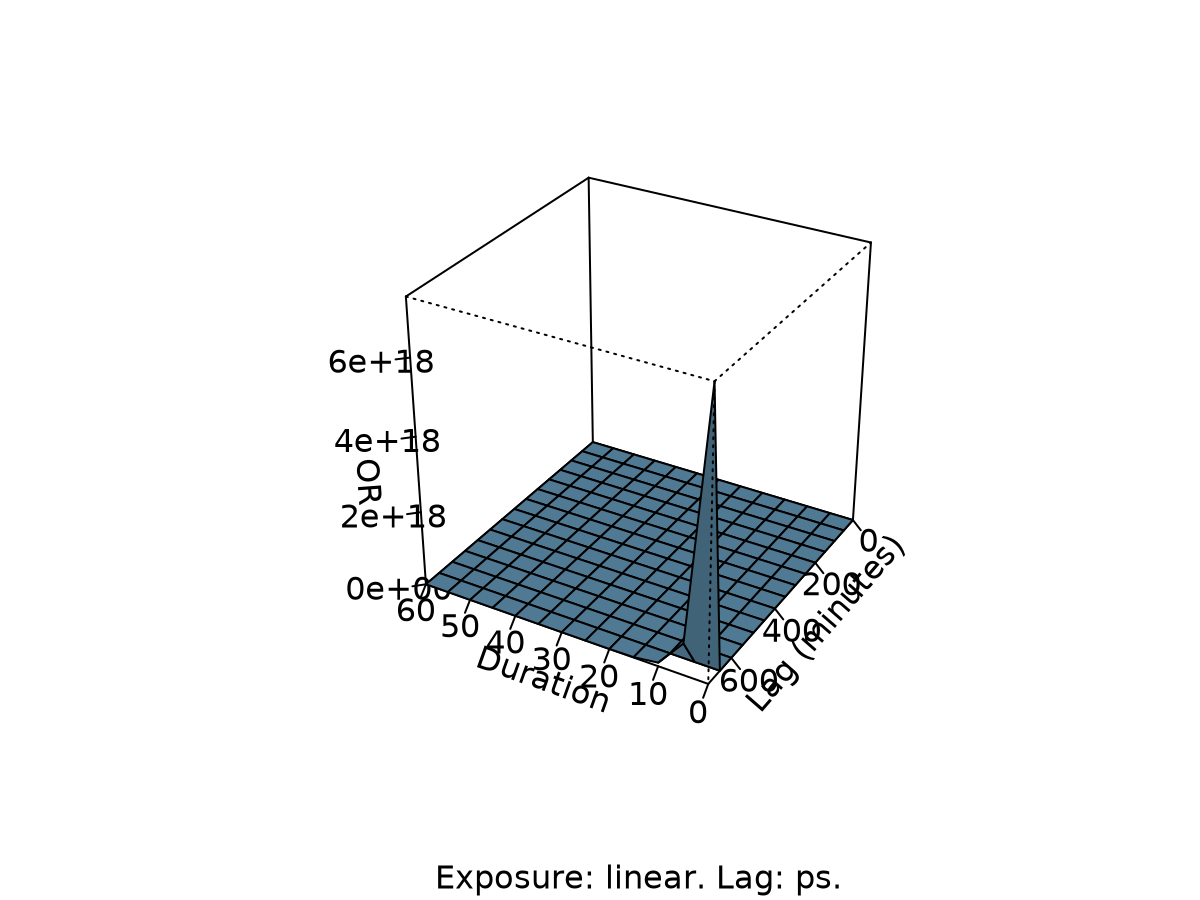}
\includegraphics[width=0.48\textwidth]{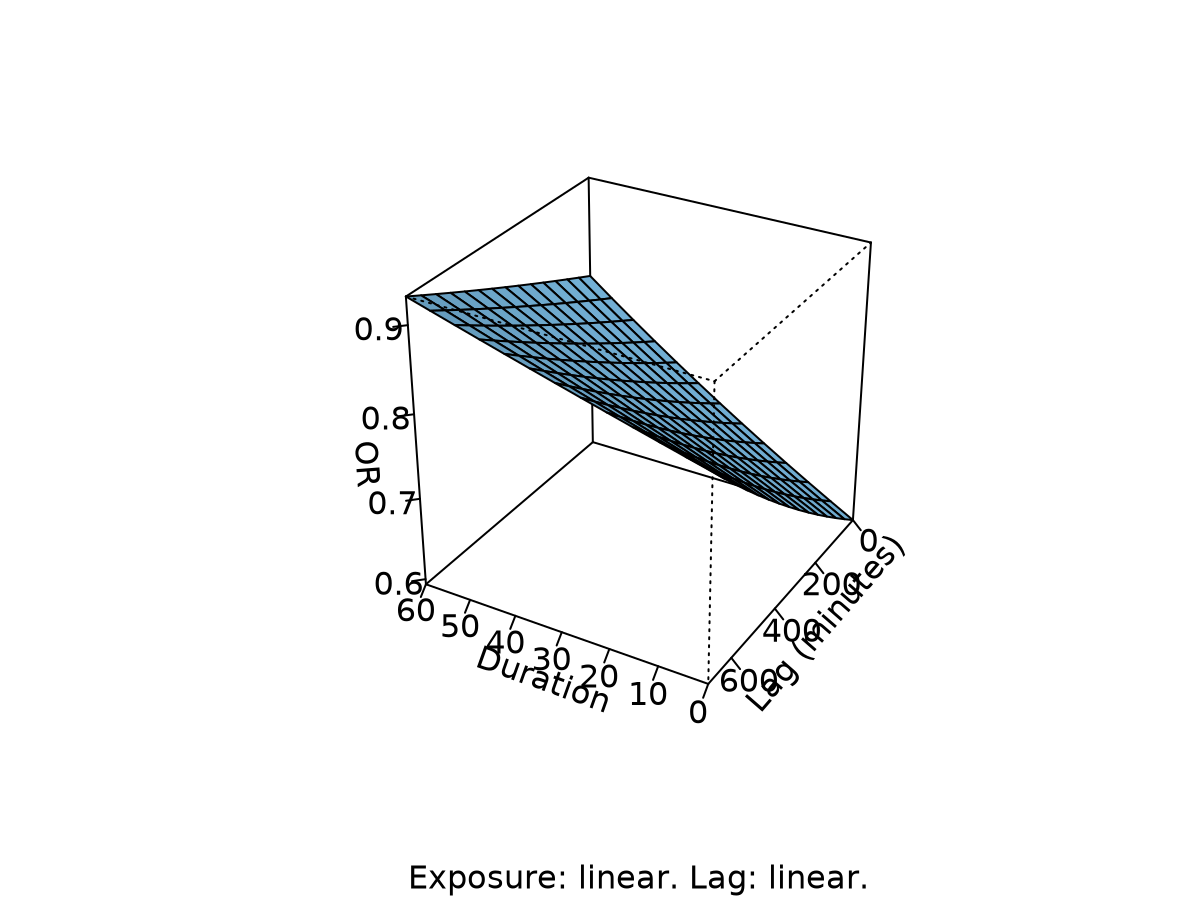}
\includegraphics[width=0.48\textwidth]{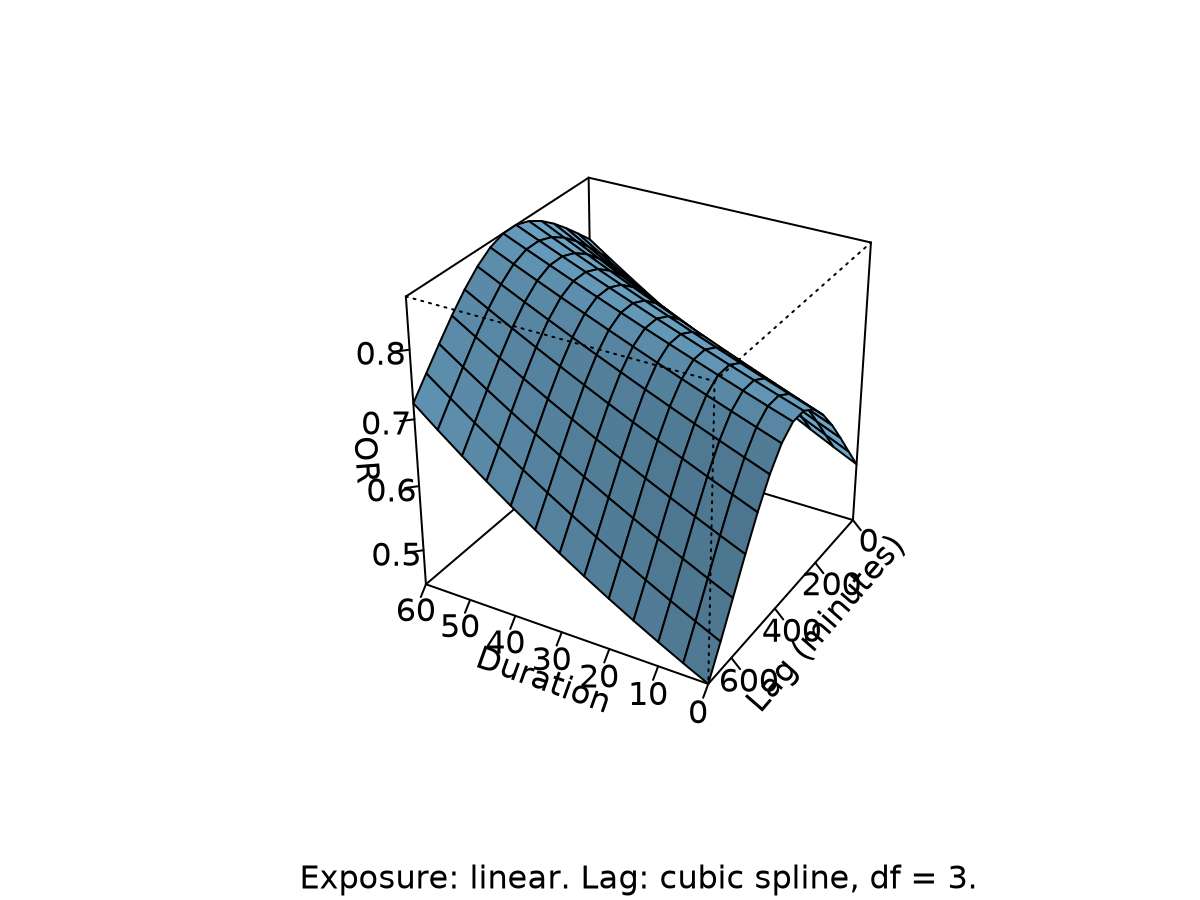}
\includegraphics[width=0.48\textwidth]{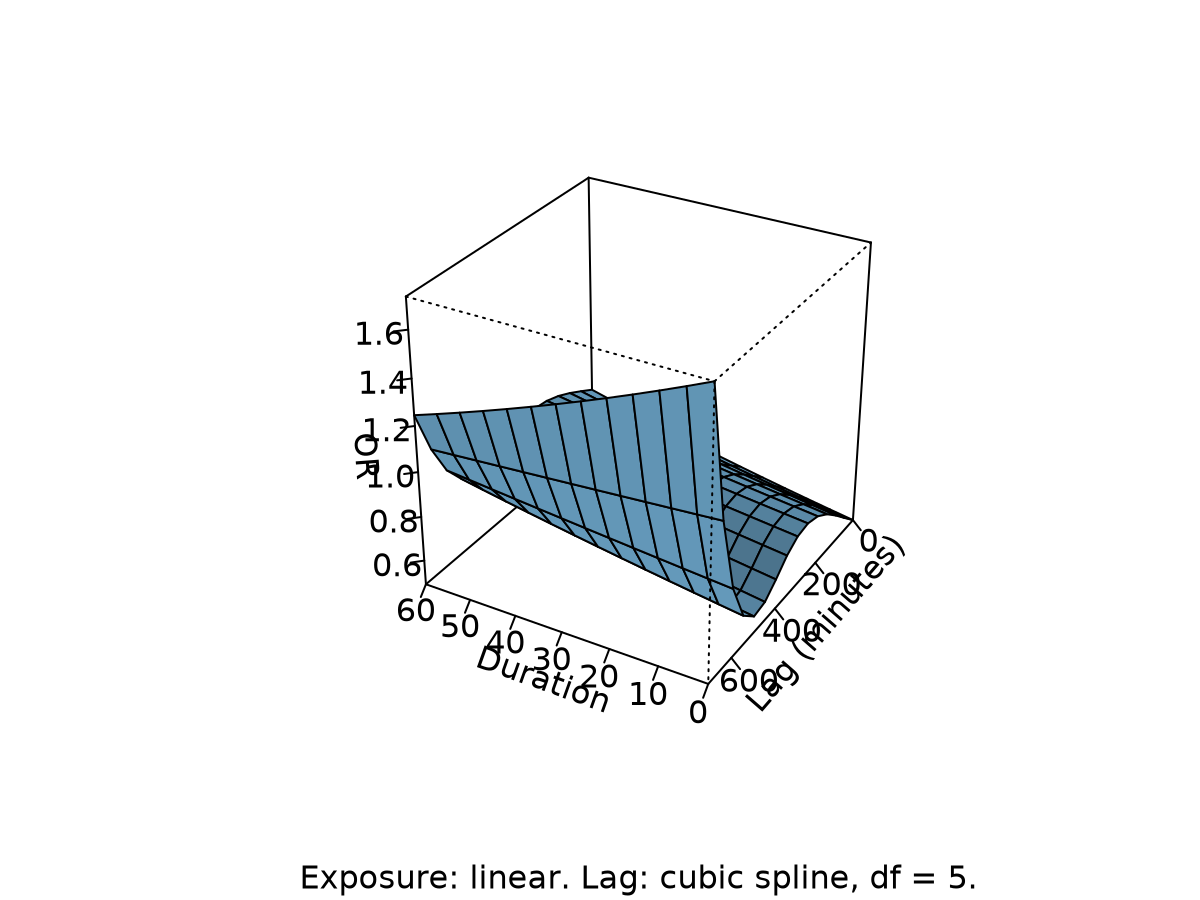}
\includegraphics[width=0.48\textwidth]{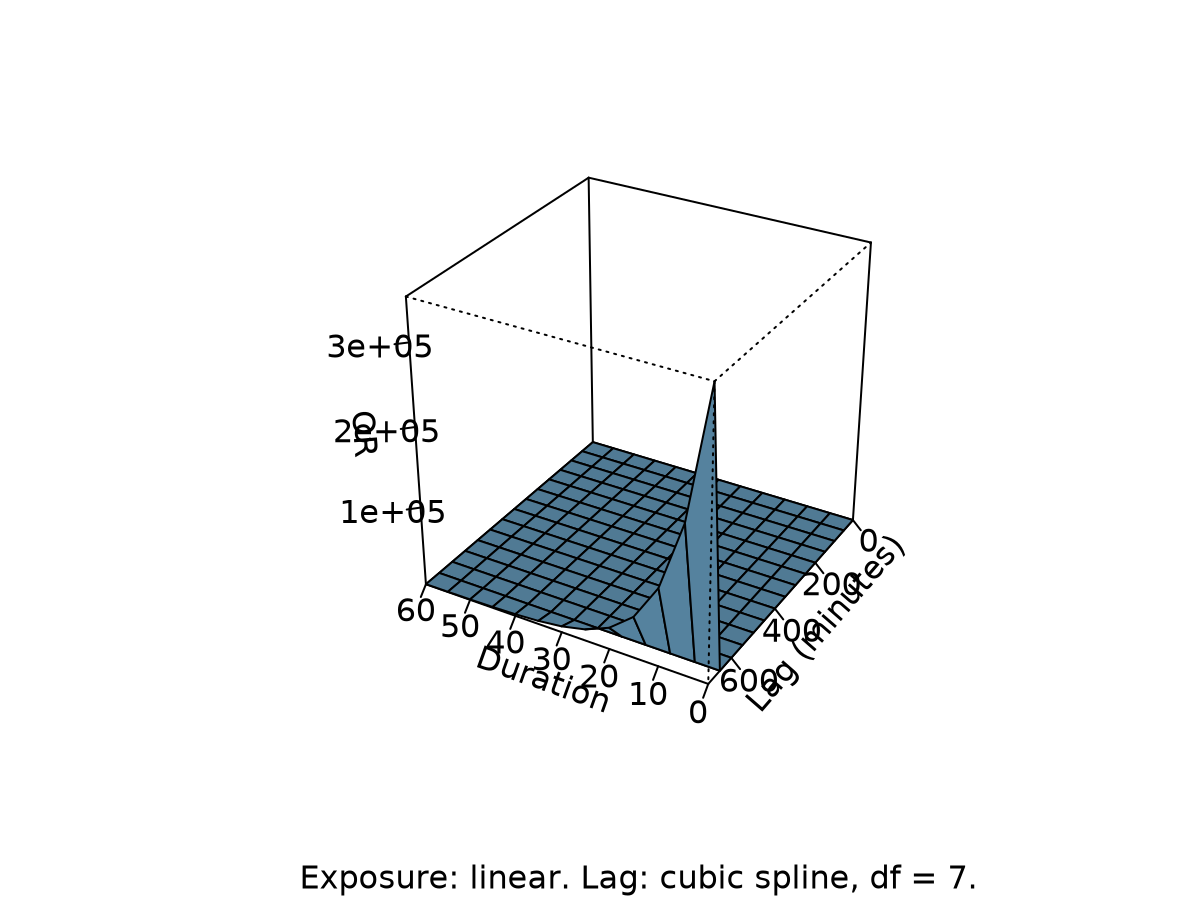}
\includegraphics[width=0.48\textwidth]{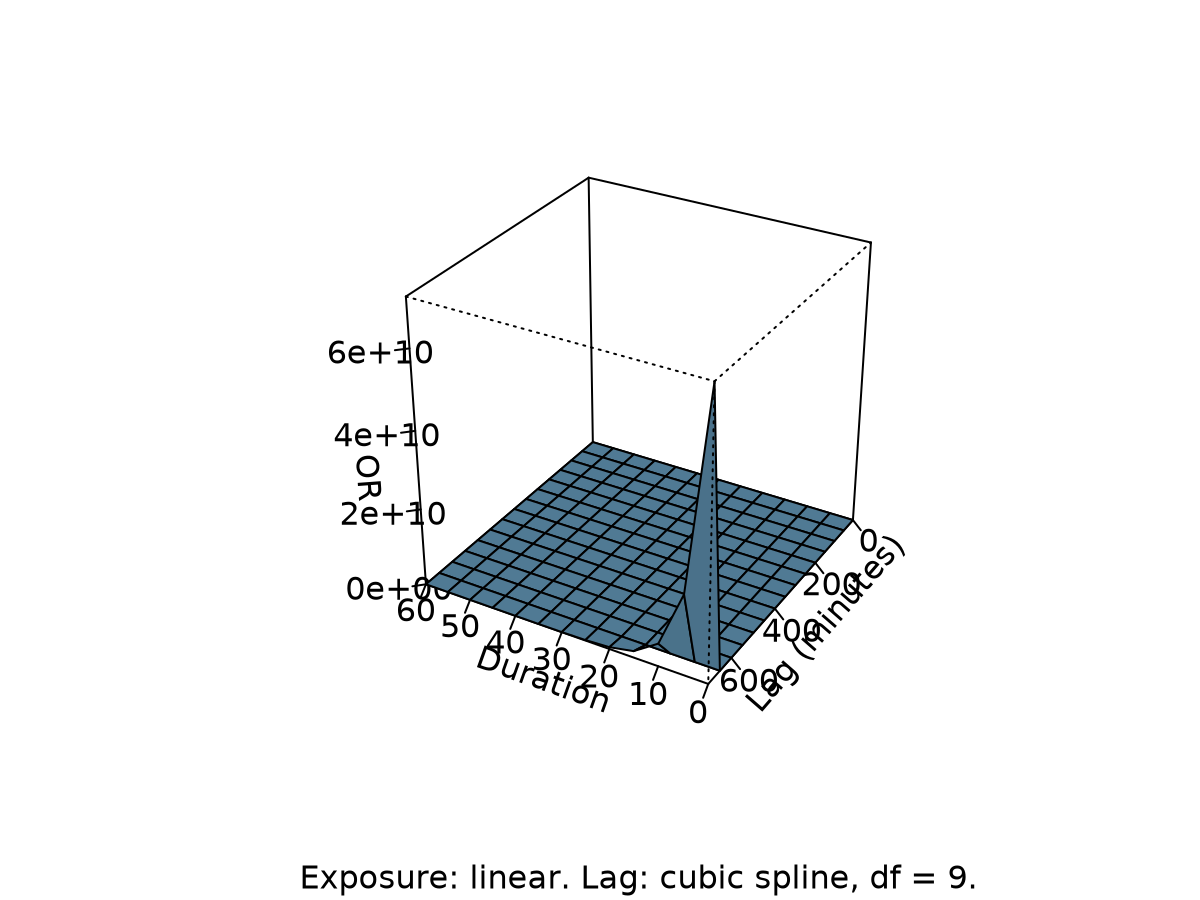}
  \caption{The estimated exposure–lag–response surface from distributed lag nonlinear models under different basis specifications. For ease of comparison, we only show the surface for duration up to 60 minutes.}
  \label{supp-fig:compare-with-dlnm}
\end{figure}

Figure \ref{supp-fig:compare-with-dlnm} presents the estimated exposure–lag–response surface. For ease of comparison, we only show the surface for duration up to 60 minutes. As discussed in Section 3.2, the interpretation of lag in this setting is not straightforward, which complicates interpretation of the estimated surface, as exposure effects must be interpreted with reference to a given lag.

Taken together, these results highlight both the flexibility of DLNM and the challenges of applying it to episodic exposure data. In particular, the need to impose a lag structure and a common temporal grid introduces modeling choices that are difficult to justify in this context and can substantially influence the results. Therefore, while DLNM is a powerful and widely used framework, it is not well aligned with the primary scientific objective of this study, namely characterizing duration-dependent risk accumulation at the episode level. 

\newpage
\bibliographystyle{apacite}
\bibliography{ref}

\end{document}